\documentclass[useAMS,usenatbib]{mn2e}

 \usepackage{times}
 \usepackage{graphicx}

\def\pcm3{{\rm\thinspace cm$^{-3}$}}

\def\contcaption{\@conttrue\SFB@caption\@captype}

\title[The Pleiades astrometric and photometric mass function]
{Astrometric and photometric initial mass functions from the UKIDSS Galactic 
Clusters Survey: I The Pleiades
\thanks{Based on observations made with the United Kingdom Infrared
Telescope, operated by the Joint Astronomy Centre on behalf of the
U.K. Particle Physics and Astronomy Research Council.}}
\author[N. Lodieu et al.]{N. Lodieu$^{1,2}$\thanks{E-mail:
nlodieu@iac.es}, N. R. Deacon$^{3}$, N. C. Hambly$^{4}$ \\
$^{1}$Instituto de Astrof\'isica de Canarias (IAC), V\'ia L\'actea s/n,
E-38205 La Laguna, Tenerife, Spain \\
$^{2}$ Departamento de Astrof\'isica, Universidad de La Laguna (ULL),
E-38205 La Laguna, Tenerife, Spain \\
$^{3}$Max-Planck-Institut f\"ur Astronomie, K\"onigstuhl 17, 69117, 
Heidelberg, Germany \\
$^{4}$Scottish Universities' Physics Alliance (SUPA),
Institute for Astronomy, School of Physics, University of Edinburgh,
Royal Observatory, Blackford Hill, \\
Edinburgh EH9 3HJ, UK} 
\begin{document}

\date{Accepted \today{}. Received \today{}; in original form \today{}}

\pagerange{\pageref{firstpage}--\pageref{lastpage}} \pubyear{2005}

\maketitle

\label{firstpage}

\begin{abstract} 
We present the results of a deep wide-field near-infrared survey of the entire
Pleiades cluster recently released as part of the UKIRT Infrared Deep Sky (UKIDSS)
Galactic Clusters Survey (GCS) Data Release 9 (DR9). We have identified a sample
of $\sim$1000 Pleiades cluster member candidates combining photometry in five
near-infrared passbands and proper motions derived from the multiple epochs
provided by the UKIDSS GCS DR9\@. We also provide revised membership for
all previously published Pleiades low-mass stars and brown dwarfs in the past
decade recovered in the UKIDSS GCS DR9 Pleiades survey based on the new 
photometry and astrometry provided by the GCS\@. 
We find no evidence of $K$-band variability in the Pleiades members 
larger than $\sim0.08$~mag. In addition, we infer a substellar binary frequency 
of 22--31\% in the 0.075--0.03 M$_{\odot}$ range for separations less than 
$\sim$100 au. We employed two independent but complementary methods to derive 
the cluster luminosity and mass functions: a probabilistic analysis and a more 
standard approach consisting of stricter astrometric and photometric cuts. We 
found that the resulting luminosity and mass functions obtained from both methods 
are very similar. We derive the Pleiades mass function in the 0.6--0.03 M$_{\odot}$ 
mass range and found that it is best reproduced by a log-normal representation
with a mean characteristic mass of $m_c=0.24^{+0.01}_{-0.03}$~M$_{\odot}$, in 
agreement with earlier studies and the extrapolation of the field mass function.
\end{abstract}


\section{Introduction}

Over the past two decades star forming regions and rich, young open 
clusters have been the focal points of numerous searches for substellar 
objects in the form of wide-field or pencial-beam surveys
\citep[e.g.][]{jameson89,hambly93,luhman99a,lucas00,bejar01,zapatero02b,tej02,moraux03,lodieu06,lodieu07a}.
Among the key milestones in the study of substellar objects, we should point
out the discovery of a first brown dwarf in a star-forming region
\citep[$\rho$\,Oph;][]{luhman97}, the identification of young L dwarfs
\citep{martin98b}, and the first young T dwarf member of $\sigma$\,Orionis
\citep[SOri\,70;][]{zapatero02b}. The common scientific driver of these surveys 
in young regions is the study of the Initial Mass Function 
\citep[IMF;][]{salpeter55,miller79,scalo86,kroupa02,chabrier05a} and its possible
universality through the imaging and spectroscopic confirmation of candidate
members in a wide variety of environments.

The UKIRT Infrared Deep Sky Survey \citep[UKIDSS;][]{lawrence07}\footnote{The
survey is described at www.ukidss.org} is a deep large-scale infrared survey.
All photometric observations, obtained with the Wide field CAMera 
\citep[WFCAM;][]{casali07} equipped with five infrared filters 
\citep[$ZYJHK$;][]{hewett06}, are pipeline-processed at the Cambridge 
Astronomical Survey Unit (CASU; Irwin et al.\ 2007, Irwin et al.\ in preparation)\footnote{The 
CASU WFCAM-dedicated  webpage can be found at http://apm15.ast.cam.ac.uk/wfcam}.
The processed data are then archived in Edinburgh and released to the user 
community through the WFCAM Science Archive \citep[WSA;][]{hambly08}\footnote{The 
WFCAM Science Archive is accessible at the URL http://surveys.roe.ac.uk/wsa}.
One of its components, the Galactic Clusters Survey (hereafter GCS) aims at
covering $\sim$1000 square degrees in 10 star-forming regions and open clusters
down to 0.03--0.01 M$_{\odot}$ (depending on the age and distance) to investigate
the universality of the initial mass function.
Each cluster is covered in $ZYJHK$ with a second epoch in $K$ to provide proper
motions with accuracies of a few milli-arcsec per year (mas/yr). The latest
GCS data release (DR) to date, DR9 on 25 October 2011, provides full
coverage of the Pleiades in those five filters along with proper motions. 
This paper is the first of a series dedicated to the astrometric and photometric 
mass functions in young and intermediate-age open clusters as well as  
star-forming regions to address the 
fundamental issue of the universality of the IMF in an homogeneous manner.

The rich Pleiades cluster has been subjected to a particularly high degree of 
scrutiny for several reasons. First, its members share a significant common
proper motion compared to neighbouring stars estimated to
($\mu_\alpha\cos\delta$, $\mu_\delta$) = ($+$19.15, $-$45.72) and
($+$20.10, $-$45.39) mas/yr by \citet{robichon99} and \citet{vanLeeuwen09}
respectively, making astrometric selection relatively straightforward.
Furthermore, the Pleiades is among the nearest well-populated open clusters
located at 134 pc from the Sun with an uncertainty of 5 pc 
\citep{johnson57,gatewood00,pinfield00,southworth05} while the
improved reduction of the Hipparcos data by \citet{vanLeeuwen09} suggests
a distance of 120.2$\pm$1.9 pc. In this work we adopt this latter estimate.
Its age has been determined using various methods, including the 
Zero-Age-Main-Sequence turn-off \citep{mermilliod81} and the lithium depletion 
boundary methods \citep{stauffer98}. The latter method yielded a most probable 
value of $125\pm8$ Myr. Moreover, reddening along the line of sight to the 
cluster is generally low, E($B-V$) = 0.03 \citep*{Odell94}.
Finally, the number of Pleiades members is large due to the numerous surveys
dedicated to the stellar and substellar components of the cluster
\citep{jameson89,hambly93,stauffer94c,zapatero97a,stauffer98b,martin98a,bouvier98,festin98,zapatero99b,hambly99,pinfield00,adams01a,moraux01,jameson02,dobbie02b,moraux03,deacon04,bihain06,bihain10a,casewell07,lodieu07c,bihain10a,casewell11}.

In this paper we present the Pleiades mass function derived from $\sim$80
square degrees surveyed in 5 passbands, one at 2 epochs, by the UKIDSS
GCS to provide photometry and astrometry for about one million sources.
These data come from the latest data release of the UKIDSS GCS, DR9\@.
In Section \ref{Pleiades:dataset} we present the photometric and astrometric
dataset employed to extract Pleiades member candidates.
In Section \ref{Pleiades:status_old_cand} we review the list of previously 
published members recovered by the UKIDSS GCS DR9 and revise their membership.
In Section \ref{Pleiades:new_cand}  we outline two methods for deriving the cluster 
luminosity function. One method relies on a relatively conservative photometric 
selection followed by the calculation of formal membership probabilities based on 
objects positions in the proper motion vector point diagram 
(Section \ref{Pleiades:new_cand_probabilistic}). The second method applies a 
rigorous astrometric selection based on the formal errors on the proper motions 
for each photometric candidate compared to the mean of the cluster 
(Section \ref{Pleiades:new_cand_phot_PM}) after more stringent multicolour cuts.
In Section \ref{Pleiades:binary} we discuss the photometric binary 
frequency in the substellar regime and compare it with previous 
estimates in the Pleiades and for ultracool field dwarfs. 
In Section \ref{Pleiades:variability} we discuss the $K$-band variability
of Pleiades cluster member candidates.
In Section \ref{Pleiades:IMF} we derive the cluster luminosity and 
(system) mass function and compare it to earlier estimates for this cluster and 
others, along with that of the field population.

%
%
\section{The astrometric and photometric dataset}
\label{Pleiades:dataset}
%

%
%
\begin{figure}
   \centering
   \includegraphics[width=\linewidth]{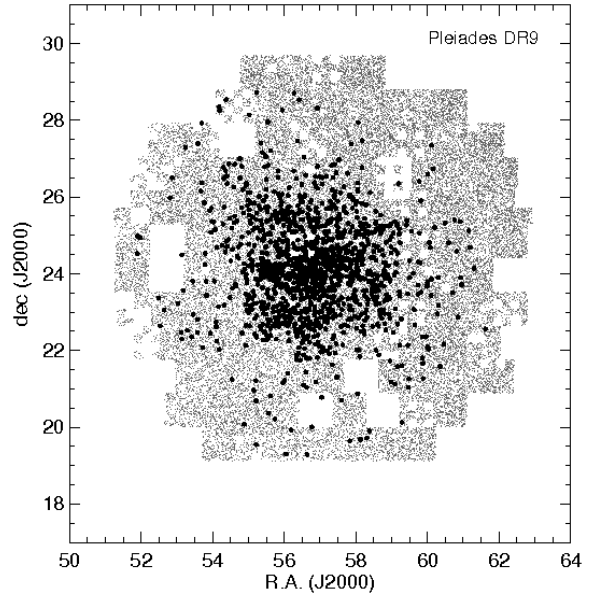}
   \caption{The coverage in the Pleiades as released by the UKIDSS GCS DR9
(grey region). The holes are due to frames removed from the GCS release due
to quality control issues. Overplotted are previously known member candidates 
recovered by the GCS DR9 (filled black dots).
}
   \label{fig_Pleiades:coverage_DR9}
\end{figure}
%

%
%
\begin{figure}
   \centering
   \includegraphics[width=\linewidth]{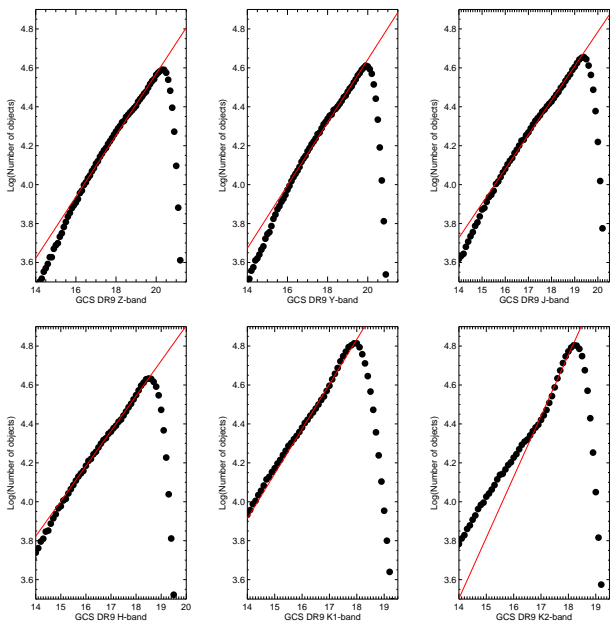}
   \caption{Completeness of the GCS DR9 dataset in the Pleiades cluster in 
each of the six filters. The polynomial fit of order 2 is shown as a red and 
defines the 100\% completeness limit of the GCS DR9 in each passband.
}
   \label{fig_Pleiades:compl_DR9}
\end{figure}
%

%
%
\subsection{Photometry}
\label{Pleiades:datasets_photometry}

We selected point sources in the full Pleiades cluster, in the region defined 
by RA=50--64 degrees and dec=18--30 degrees. We retrieved the catalogue using a 
Structure Query Language (SQL) query similar to our earlier study of the Pleiades 
\citep{lodieu07c} and applied it to the DR9 release of the GCS (see also the 
SQL query in appendix A of \cite{lodieu07a}).

\begin{figure}
   \includegraphics[width=1.00\linewidth]{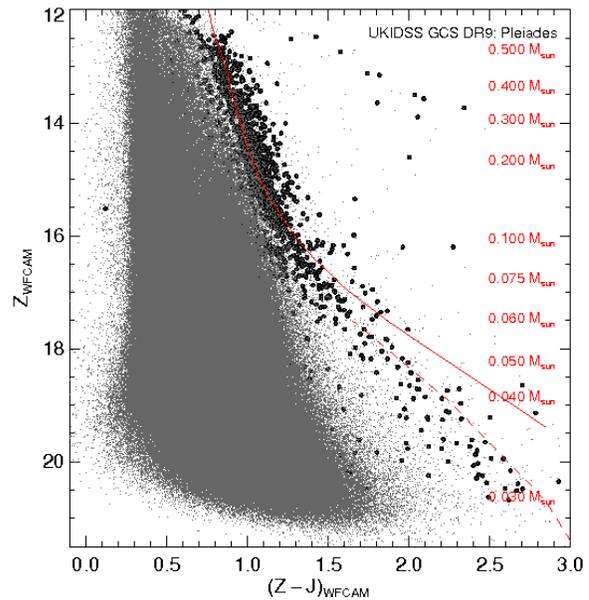}
   \caption{($Z-J$,$Z$) CMD for $\sim$80 square degrees in the Pleiades extracted
from the UKIDSS Galactic Cluster Survey Data Release 9\@. Previously published
Pleiades member candidates are overplotted as filled dots. The mass scale is shown 
on the right hand side of the diagram and extends down to $\sim$0.02 M$_{\odot}$, 
according to the NextGen (solid line) and DUSTY (dashed line)
models \citep{baraffe98,chabrier00c}.
}
   \label{fig_Pleiades:ZJZcmd_alone_DR9}
\end{figure}

Briefly, we selected only good quality point sources in all passbands
and included $Z$ and $Y$ non detections to allow for cooler brown dwarf candidates 
to be extracted. We did not impose a detection in the second $K$-band either because the 
proper motions are not only computed with the two $K$-band epochs but with any 
filter observed at a different time (see Section \ref{Pleiades:dataset_PM}).
We limited our selection to sources fainter than $Z$ = 11.3, $Y$ = 11.5, 
$J$ = 11.0, $H$ = 11.3, and $K$1 = 9.9 mag where the GCS saturates.
The completeness limits, taken as the magnitude where the straight line
fitting the shape of the number of sources as a function of magnitudes falls
off, are $Z$ = 20.2, $Y$ = 19.8, $J$ = 19.3, $H$ = 18.4, $K$1 = 17.9, and
$K$2 = 18.1 mag (Fig.\ \ref{fig_Pleiades:compl_DR9}).
The SQL query includes a cross-match with the Two Micron All Sky Survey
\citep[2MASS;][]{cutri03,skrutskie06} if available.
The main change here compared to our previous study of the Pleiades made with 
the GCS DR1, apart from the significant increase in areal coverage,
is the inclusion of the proper motions determined from the multiple 
passband coverage taken at different epochs by the GCS and released
in DR9 (see Section \ref{Pleiades:dataset_PM} for more details).
The query returned a total of 937,723 sources over $\sim$80 square degrees. 
The full coverage is displayed in Fig.\ \ref{fig_Pleiades:coverage_DR9} 
and the resulting ($Z-J$,$Z$) colour-magnitude diagram is shown in 
Fig.\ \ref{fig_Pleiades:ZJZcmd_alone_DR9}.
We note that the theoretical isochrones plotted in the different figures of
the paper were specifically computed for the WFCAM set of filters and kindly
provided by I.\ Baraffe and F.\ Allard.

%
%

%
%
\subsection{Astrometry}
\label{Pleiades:dataset_PM}

Proper motion measurements are available in the WFCAM Science Archive for UKIDSS
data releases from DR9 for all the wide/shallow surveys with multiple epoch
coverage in each field (i.e.\ the LAS, GCS and GPS). Details of the procedure will 
appear elsewhere (Collins \& Hambly 2012), but here we give a brief 
description of the process. 

When complete, each field imaged in the UKIDSS surveys is covered by a set of 
detector frames in various passbands with one passband revisited at least once. 
In general, these frames may have been taken at any time over the lifetime of 
the survey (at the time of writing, $\sim6$~years) resulting in multiple epoch 
coverage for all sources. The approach taken in the WSA for computing proper 
motions is simple\footnote{we emphasise that this work was done within the WSA 
for all the UKIDSS wide/shallow surveys, and not carried out for the sake of 
this paper only}: first a set of local plane coordinates was established for all 
detections in all frames that have any systematic offsets in absolute positions 
minimised; then proper motions were measured from linear least--squares fits in 
these local plane coordinates as a function of time. In the first process, for 
each set of frames, one reference frame was taken to map each other `slave' frame 
onto that master using {\em all} available detections and a simple linear `plate' 
model. Working in local plane coordinates, the model comprises 6 coefficients 
allowing for independent zeropoint shifts and scale changes in both coordinates, 
rotation and non--orthogonality between the coordinate axes (shear). The reason 
for applying these local models is that the absolute astrometry of each frame is 
done with respect to relatively bright 2MASS standard stars, so the zeropoint of 
any proper motions derived by simply taking these raw positions would be defined 
by these relatively nearby stars, and would be subject to any bulk motions and/or 
drift of such stars as seen from our vantage point in space. The idea was to 
define a zeropoint for the proper motions across all surveys that is as close as 
possible to a true zero, i.e.~one in which the average galaxy and quasar proper 
motions would be zero. Clearly this is not possible in low--latitude sightlines 
and/or for relatively shallow detection lists containing few identified 
extragalactic sources, so the best that can be done is to use the faintest and 
hence on average most distant stars possible. Since the number counts are 
dominated by the fainter stars, all stars were simply used. Note that rather 
than weighting these local model fits by the formal errors on each detection, 
unit--weight fits were chosen. Again, this is because the brighter, and on 
average nearer stars would carry the most weight given their low centroiding 
errors, but the local mapping models should not be biased towards the possibly 
drifting reference frame that would be defined by such stars which exhibit 
significant angular motions simply because of their proximity to the Sun.

Once a set of mapped local plane coordinates was set for every source paired 
across the detections available from the set of frames, a weighted linear 
least--squares fit is done for each coordinate as a function of time, resulting 
in four astrometric parameters (coordinates at a reference epoch along with 
proper motions in those coordinates) plus formal errors and a standard 
goodness--of--fit  parameter (a reduced chi--squared statistic).

%
%
\section{Cross-match with previous surveys}
\label{Pleiades:status_old_cand}

We compiled a list of Pleiades member candidates published over the past
decades by various groups (Table \ref{tab_Pleiades:early_summary}) to update their 
membership status with the photometry 
and astrometry in DR9 of the GCS (Table \ref{tab_Pleiades:early_DR9}). 
This list will serve as starting point to identify new Pleiades members in the 
GCS data and derive the cluster luminosity and mass functions. 

For the brightest members, we used the extensive compilation of 1417 sources 
from \citet{stauffer07} which includes candidate cluster members from several 
large-scale proper motion studies of the Pleiades
\citep{trumpler21,hertzsprung47,artyukhina69,jones81,haro82,vanLeeuwen86,stauffer91,hambly93,pinfield00,adams01a,deacon04}.
We added candidates from a large number of additional papers dedicated to 
the Pleiades low-mass stars and brown dwarfs over the past 20 years, many sources
being common to several studies which surveyed independently the same region
of the cluster. The references are listed in 
Table \ref{tab_Pleiades:early_summary} along with the original numbers of
sources published by each study (All) and the corresponding numbers of candidates 
covered by the GCS (DR9). The success rate in the recovery of published candidate 
members is usually quite high (Table \ref{tab_Pleiades:early_summary}) because 
the GCS covers now the entire Pleiades 
cluster, as can be seen from Fig.\ \ref{fig_Pleiades:coverage_DR9}. Earlier
studies such as \cite{hambly93}, \citet{adams01a}\footnote{We should mention that
this catalogue was not publish but is included in the \citet{stauffer07}
compilation}, and the list of 916 high-probability member candidates from
\citet{deacon04} are contained in \citet{stauffer07} fnd ocused 
not only on Pleiades very-low-mass stars but also on brighter members which are 
saturated on the UKIDSS images. Thus these higher mass members are not retrieved 
by our SQL query due to our photometric cuts at the bright end of the survey.
Similarly, many L and T dwarf candidates reported by \citet{bihain06} and
\citet{casewell07}\footnote{Four objects have wrong coordinates, rectified in
the erratum of this paper \citep{casewell10}} are too faint to be detected on 
the GCS images. Some of the earlier Pleiades candidates are not 
recovered mainly because (Table \ref{tab_Pleiades:early_summary}):

\begin{itemize}
\item 341 sources brighter than our saturation limits equivalent to 58.5\% of 
previously published sources not recovered by our SQL query (mainly coming from
early surveys as mentioned above)
\item 185 sources (i.e.\ 31.7\%) missing images in $J$, $H$, or $K$1
\item 10 sources located in holes of the UKIDSS GCS coverage due to quality control
issues (1.7\% of all published members)
\item 8 (or 1.37\%) known Pleiades candidates located beyond our 3 arcsec 
cross-match limit, possibly because that they are higher proper motion non-members
\end{itemize}

To quantify the completeness limit of the GCS Pleiades dataset, we have
listed in Column 1 of Table \ref{tab_Pleiades:early_summary} (numbers in
brackets) the numbers of previously-published candidates within the magnitude
range probed by the GCS (see Section \ref{Pleiades:datasets_photometry}).
The average of the percentages listed in the last column of 
Table \ref{tab_Pleiades:early_summary} amounts for 92.2\%, not taking into
account the sample of \citet{zapatero97a} due to problems with their coordinates.
We also considered the most complete and updated sample of high-mass stars and 
low-mass Pleiades member candidates published by \citet{stauffer07}. We recovered
94\% of their sources, suggesting that overall our GCS sample is at least 
92--94\% complete.

Table \ref{tab_Pleiades:early_DR9} is provided in electronic format only 
with a total of 3196 known Pleiades member candidates reported in the literature.
From this sample we removed the multiple detections and kept all different
names from earlier studies in the last column for future searches: we are left
with 1379 Pleiades candidates. We give the GCS DR9 coordinates of these 1379
Pleiades candidates,
the $ZYJHK$1$K$2 photometry, the proper motions in right ascension and declination
with their respective errors as well as the $\chi^{2}$ value which represents
the reduced chi-squared statistic of the astrometric fit for each source. This
parameter is equal to the usual chi-squared statistic (sum of normalised 
residuals) divided by the degrees of freedom (i.e.\ number of data points minus 
the number of fitted parameters). The penultimate column supplies 
the membership probability for 1067 of the 1379 previously reported Pleiades
candidates (see Section \ref{Pleiades:new_cand_probabilistic_method} for the 
method). A total of 312 out of 1379 Pleiades candidates have no membership 
probabilities because they are not classified as Pleiades members by the 
probabilistic selection (see Section \ref{Pleiades:new_cand_probabilistic} for 
the photometric and astrometric criteria). They are divided into four groups 
as follows: 190 sources detected in $ZYJHK$ but classified as proper
motion non members, 42 objects without $Z+Y$ photometry, 74 objects 
without $Z$ only, and 6 sources without $Y$ only.
The last column gives the old names used by earlier studies
(names from different authors are separated by an underscore ``\_'').

Previously-published Pleiades member candidates not recovered in the GCS DR9
are listed in Table \ref{tab_Pleiades:known_NOT_in_GCSDR9} with their coordinates
and old names from earlier studies. After removing common sources, we are left
with 544 known Pleiades member candidates not in our sample. The large majority
of these sources are either too bright or too faint to in the GCS database or 
photometric and/or astrometric non members of the Pleiades.

%
%
\begin{table}
  \caption{Updated membership of Pleiades member candidates published
in the literature and recovered by the GCS DR9\@. Papers dedicated to the
Pleiades over the past two decades are ordered by year.
References are: \citet{hambly93}, \citet{zapatero97a}, \citet{festin98},
\citet{bouvier98}, \citet{zapatero99b}, \citet{hambly99}, \citet{pinfield00},
\citet{moraux01}, \citet{jameson02}, \citet{dobbie02b}, \citet{moraux03},
\citet{deacon04}, \citet{bihain06}, \citet{lodieu07c}, \citet{casewell07}, and 
\citet{stauffer07}. Columns 2 and 3 give the numbers of sources published by the 
reference given in Column 1 and the numbers of sources recovered in GCS DR9, 
respectively. Numbers in brackets in Column 1 represent all sources in a 
given catalogue within the magnitude range probed by the GCS\@.
Columns 4 and 5 give the numbers of high-probability members (p$\geq$60\%) and
non members (NM) according to our probabilistic approach (first number) and
method \#2 (second number). The last column gives the percentages of sources 
recovered in the GCS DR9 (i.e.\ Column 2 divided by numbers in brackets in
Column 1)\@. 
}
  \label{tab_Pleiades:early_summary}
  \begin{tabular}{l c c c c r}
  \hline
Survey        &  All  & DR9 & Memb  &  NM  & \%   \cr
  \hline
Hambly1993    &  440 (440) & 418 &  303/302    &  115/116  &   95.0  \cr
Zapatero1997a &    9  (9)  &   2 &   1/1       &    1/1    &   22.2  \cr
Festin1998    &   45  (44) &  37 &   24/24     &   13/13   &   84.1  \cr
Stauffer1998  &   20  (19) &  16 &   11/10     &    5/6    &   84.2  \cr
Bouvier1998   &   26  (26) &  25 &   12/15     &   13/10   &   96.2  \cr
Zapatero1999  &   46  (44) &  38 &    8/13     &   30/25   &   86.4  \cr
Hambly1999    &    9   (9) &   9 &    5/6      &    4/3    &  100.0  \cr
Pinfield2000  &  339 (338) & 320 &  185/187    &  135/133  &   94.7  \cr
Moraux2001    &   25  (25) &  25 &   12/15     &   13/10   &  100.0  \cr
Dobbie2002    &   90  (87) &  61 &    8/9      &   53/52   &   70.0  \cr
Moraux2003    &  109 (108) & 107 &   74/74     &   33/33   &   99.1  \cr
Deacon2004    &  916 (746) & 674 &  467/450    &  207/224  &   90.3  \cr
Bihain2006    &   34  (31) &  28 &   11/14     &   17/14   &   90.3  \cr
Lodieu2007    &  456 (456) & 454 &  376/376    &   78/78   &   99.6  \cr
Casewell2007  &   23  (16) &  16 &    4/7      &    9/6    &  100.0  \cr
Stauffer2007  & 1416 (944) & 888 &  567/639    &  321/249  &   94.0  \cr
 \hline
\end{tabular}
\end{table}

%
%
%
\section{New substellar members in the Pleiades}
\label{Pleiades:new_cand}
\subsection{Probabilistic approach}
\label{Pleiades:new_cand_probabilistic}
\subsubsection{Method}
\label{Pleiades:new_cand_probabilistic_method}

In this section we outline the probabilistic approach we employed to select 
low-mass stars and brown dwarf member candidates of the Pleiades using photometry 
and astrometry from the UKIDSS GCS DR9\@. This method is described in detail
in \citet{deacon04} and \citet{lodieu07c}. The main steps are:
\begin{enumerate}
\item Define the cluster sequence using candidates published in the literature
within the area covered by the latest release of the GCS
\item Make a conservative photometric cut in the ($Z-J$,$Z$) diagram to 
include known members and identify new cluster member candidates (dashed 
lines on the top-left panel in Fig.\ \ref{fig_Pleiades:YJK_cmds_method2}).
\item Analyse the vector point diagram in a probabilistic manner to assign 
a membership probability for each photometric candidate with a proper motion 
measurement (Section \ref{Pleiades:new_cand_probabilistic_proba}).
\item Obtain an illustrative list of high probability cluster members by choosing a specific 
threshold for the membership, chosen as p$\ge$0.6 here
\item Derive the luminosity and mass function using all candidates with membership
probabilities without any threshold for a complete count of the membership
\end{enumerate}

%
%
\begin{figure}
   \includegraphics[width=\linewidth]{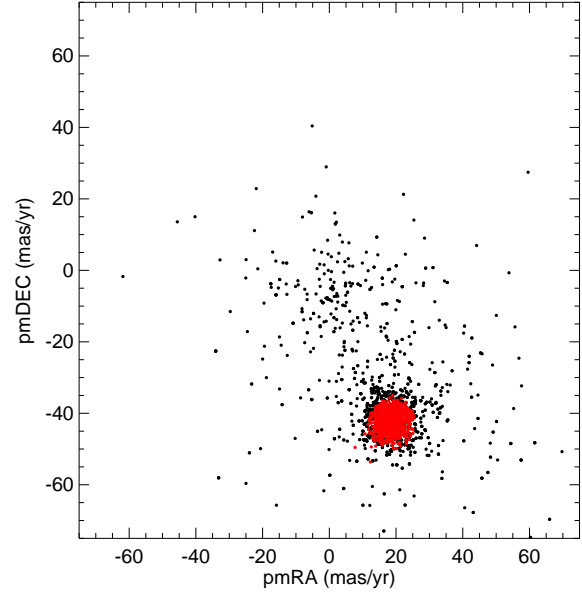}
   \caption{Vector point diagram showing the proper motion in right ascension
(x-axis) and declination (y-axis) for previously-known member candidates
recovered by the GCS DR9 (filled black dots) and the new member candidates
with membership probabilities higher than 60\% identified in this work (red dots).
}
   \label{fig_Pleiades:diagram_VPD}
\end{figure}
%

%
%
\subsubsection{Membership probabilities}
\label{Pleiades:new_cand_probabilistic_proba}

In order to calculate formal membership probabilities we used the same 
technique as \citet{deacon04} and \citet{lodieu07c} to fit distribution functions 
to proper motion vector point diagrams \citep{hambly95}. We refer the reader
to those papers for more details and additional equations.
First we have rotated the vector point diagram so the cluster lies 
on the y-axis using the rotation transformation below 
(Equations \ref{eq:rot1} and \ref{eq:rot2}):
\begin{equation}
\mu_{x1} = 0.03896 \times \mu_{x} - 0.921 \times \mu_{y}
\label{eq:rot1}
\end{equation}
\begin{equation}
\mu_{y1} = 0.03896 \times \mu_{y} - 0.921 \times \mu_{x}
\label{eq:rot2}
\end{equation}
corresponding a rotation angle of 23.7 degrees, assuming a relative proper 
motion of (19.7,$-$44.82) mas/yr for the Pleiades as measured on the vector
point diagram created from the GCS DR9 data \citep[slightly different from the
Hipparcos absolute motion;][]{vanLeeuwen09}. It is common to refer to the proper 
motions in the rotated vector point diagram as $\mu_{x1}$ and $\mu_{y1}$.
 
We have assumed that there are two contributions to the total 
distribution $\phi(\mu_{x},\mu_{y})$, one from the cluster 
($\phi_{c}(\mu_{x},\mu_{y})$) and one from the field stars
($\phi_{f}(\mu_{x},\mu_{y})$). The fitting region was delineated
by $-$50 $< \mu_{x} <$ 50 mas/yr and 20 $< \mu_{y} <$ 70 mas/yr.
These were added by means of a field star fraction $f$ to 
yield an expression for $\phi$ given in Equation \ref{eq:phi}:
\begin{equation}
\phi(\mu_{x},\mu_{y}) = f \phi_{f}(\mu_{x},\mu_{y}) + (1-f) \phi_{c}(\mu_{x},\mu_{y})
\label{eq:phi}
\end{equation}

We have assumed that the cluster distribution is characterised by
a bivariate Gaussian with a single standard deviation 
$\sigma$ and mean proper motion values in each axis 
$\mu_{xc}$ and $\mu_{yc}$ (Equation \ref{eq:phic}):

\begin{equation}
\phi_{c} \propto exp \left(-\frac{(\mu_{x} - \mu_{xc})^{2}+(\mu_{y} - \mu_{yc})^{2}}{2 \sigma^{2}}\right) 
\label{eq:phic}
\end{equation}

The field star distribution was fitted by a single Gaussian in
the $x$ axis (with standard deviation $\Sigma_{x}$ and mean 
$\mu_{xf}$) and a declining exponential in the $y$ axis with 
a scale length $\tau$.
The use of a declining exponential is a standard method
\citep[e.g.][]{jones91} and is justified in that the field star 
distribution is not simply a circularly-symmetric error 
distribution (i.e.\ capable of being modelled as a 2d 
Gaussian) - rather there is a prefered direction of
real field star motions resulting in a characteristic velocity
ellipsoidal signature, i.e.\ a non-Gaussian tail, in the 
vector point diagram. This is best modelled (away from the 
central error-dominated distribution) as an exponential in the 
direction of the antapex (of the solar motion).

The best fitting set of parameters were chosen using a maximum likelihood 
method \citep[see][]{deacon04}. However in a deviation from this method we 
did not fit for the standard deviation of the cluster proper motions ($\sigma$). 
Instead we calculated the mean astrometric error for all objects in each 
magnitude range and used this as our cluster standard deviation.
This fitting process was tested by \citet{deacon04} where
simulated data sets were created and run through the fitting process
to recover the input parameters. These tests produced no significant 
offsets in the parameter values \citep[see Table 3 and Appendix A 
of][for results and more details on the procedure]{deacon04}.
Hence, we have calculated the formal membership probabilities as,

\begin{equation}
p=\frac{\phi_{c}}{f\phi_{f}+(1-f)\phi_{c}}
\end{equation}
%
%
\begin{table}
  \caption{Summary of the results after running the programme
to derive membership probabilities. For each $Z$ magnitude range,
we list the number of stars used in the fit (Nb), the field star 
fraction f, and parameters describing the cluster and field star 
distribution. Units are in mas/yr except for the number of 
stars and the field star fraction f. The cluster star distribution
is described by the mean proper motions in the x and y 
directions ($\mu_{x_{c}}$ and $\mu_{y_{c}}$) and a standard
deviation $\sigma$. Similarly, the field star distribution is
characterised by a scale length for the y axis ($\tau$), a
standard deviation $\Sigma_{x}$, and a mean proper motion
in the x direction ($\mu_{x_{f}}$).
}
  \label{tab_Pleiades:prob_results}
  \begin{tabular}{@{\hspace{0mm}}c c c c c c c c c@{\hspace{0mm}}}
  \hline
$Z$ & Nb & f & $\sigma$ & $\mu_{x_{c}}$ & $\mu_{y_{c}}$ & $\tau$ & $\Sigma_{x}$& $\mu_{x_{f}}$ \cr
  \hline
12--13 &  486 & 0.86 & 3.22 & -1.36 & 44.65 & 15.51 & 17.97 & -9.48 \\ 
13--14 &  853 & 0.79 & 3.14 &  0.21 & 46.25 & 14.95 & 17.46 & -9.22 \\ 
14--15 & 1394 & 0.78 & 3.21 & -0.11 & 46.12 & 13.50 & 16.95 & -9.13 \\ 
15--16 & 1712 & 0.86 & 3.28 &  0.47 & 45.60 & 13.19 & 16.06 & -9.45 \\ 
16--17 & 1885 & 0.95 & 3.41 &  0.68 & 45.25 & 11.16 & 15.16 & -9.79 \\ 
17--18 & 1256 & 0.96 & 3.70 & -0.26 & 44.99 &  9.58 & 14.28 & -10.34 \\ 
18--19 &  493 & 0.93 & 4.28 & -1.23 & 44.07 & 11.02 & 15.09 & -11.18 \\ 
19--20 &  358 & 0.92 & 5.93 & -0.63 & 45.76 & 10.30 & 15.72 & -9.86 \\ 
20--21 &  369 & 0.84 & 9.62 &  7.80 & 41.75 & 10.80 & 17.49 & -7.76 \\ 
 \hline
\end{tabular}
\end{table}

As the astrometric errors are a function of magnitude we split
the sample into nine bins. Each bin was one magnitude wide and the constituent 
stars used to fit for six of the seven parameters in the same way as described 
in \citet{deacon04}. We used bins of one magnitude to have a sufficient number
of Pleiades members in each range. In the faintest bins where the astrometric
errors increase rapidly, the number of cluster stars was so small that we fixed 
the location of the cluster on the vector point diagram ($\mu_{xc}$ and $\mu_{yc}$) 
to the values from a brighter bin. The other parameters were fitted as normal. 
A summary of the fitted parameters from the probabilistic analysis described
above is given in Table \ref{tab_Pleiades:prob_results}.

\citet{deacon04} test the reliability of this method in Section 2.6 and find 
that the method accurately recovers the parameters from simulated datasets. 
We refer the reader to Table 3 in this paper for approximate errors on 
parameters but note that these tests were performed on simulated
datasets with a much larger fraction of cluster stars. Hence our cluster
parameters will likely have higher errors than those quoted. Some parameters
such as $\tau$ and $\sigma_{x}$ vary between magnitude bins due to noise
as a result of two competing factors. As we go fainter the real proper motions 
of our field star contaminants decrease as our typical field star will be more 
distant but the measurement errors on the proper motions increase. Hence these 
two parameters which describe the proper motions of the field stars initially 
fall with increasing magnitude but as the measurement errors (as traced by the 
parameter $\sigma$) blow up, these in turn also increase.

%
%
\subsubsection{Probabilistic sample}
\label{Pleiades:new_cand_probabilistic_phot}

\begin{figure*}
   \centering
   \includegraphics[width=0.49\linewidth]{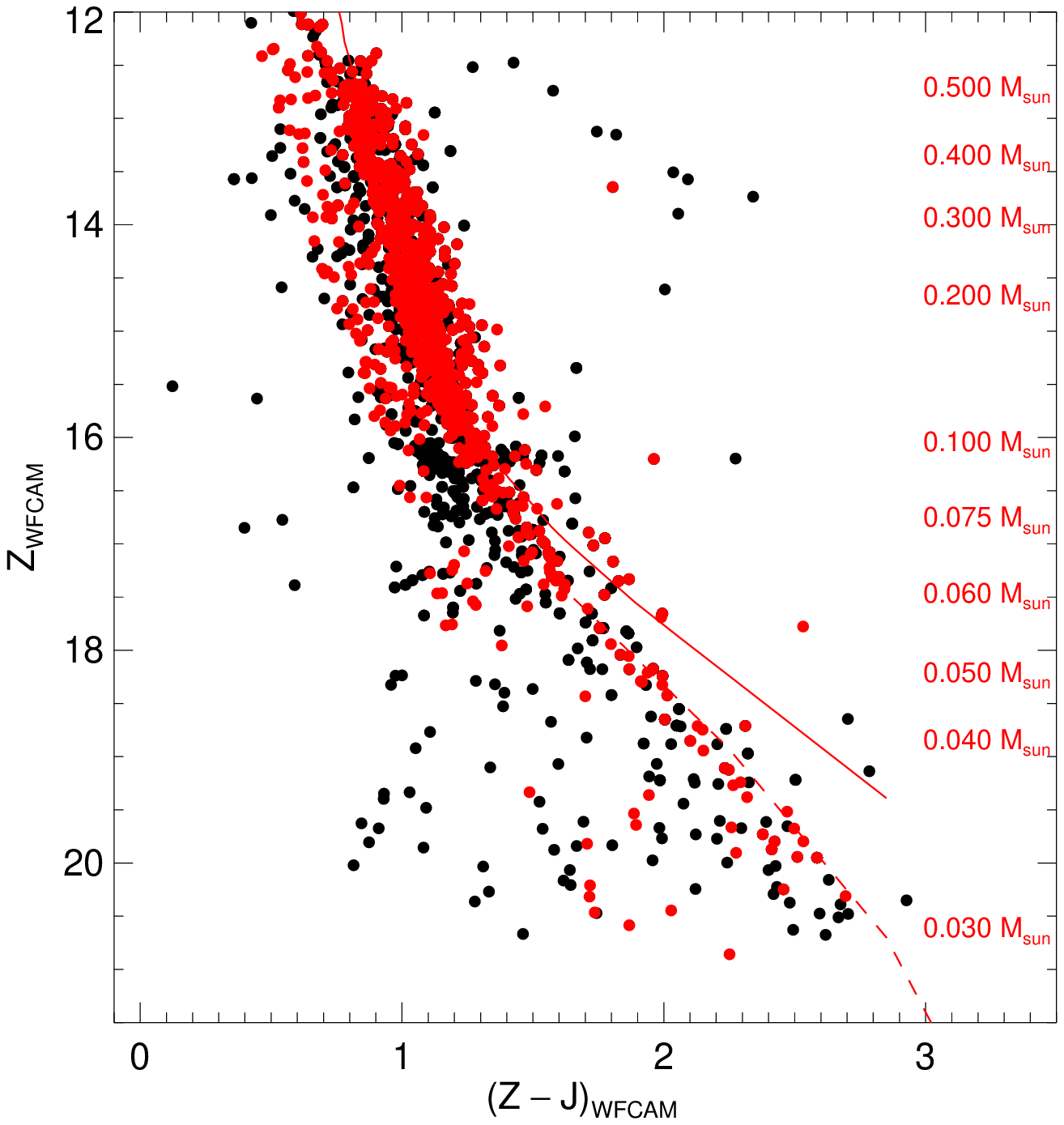}
   \includegraphics[width=0.49\linewidth]{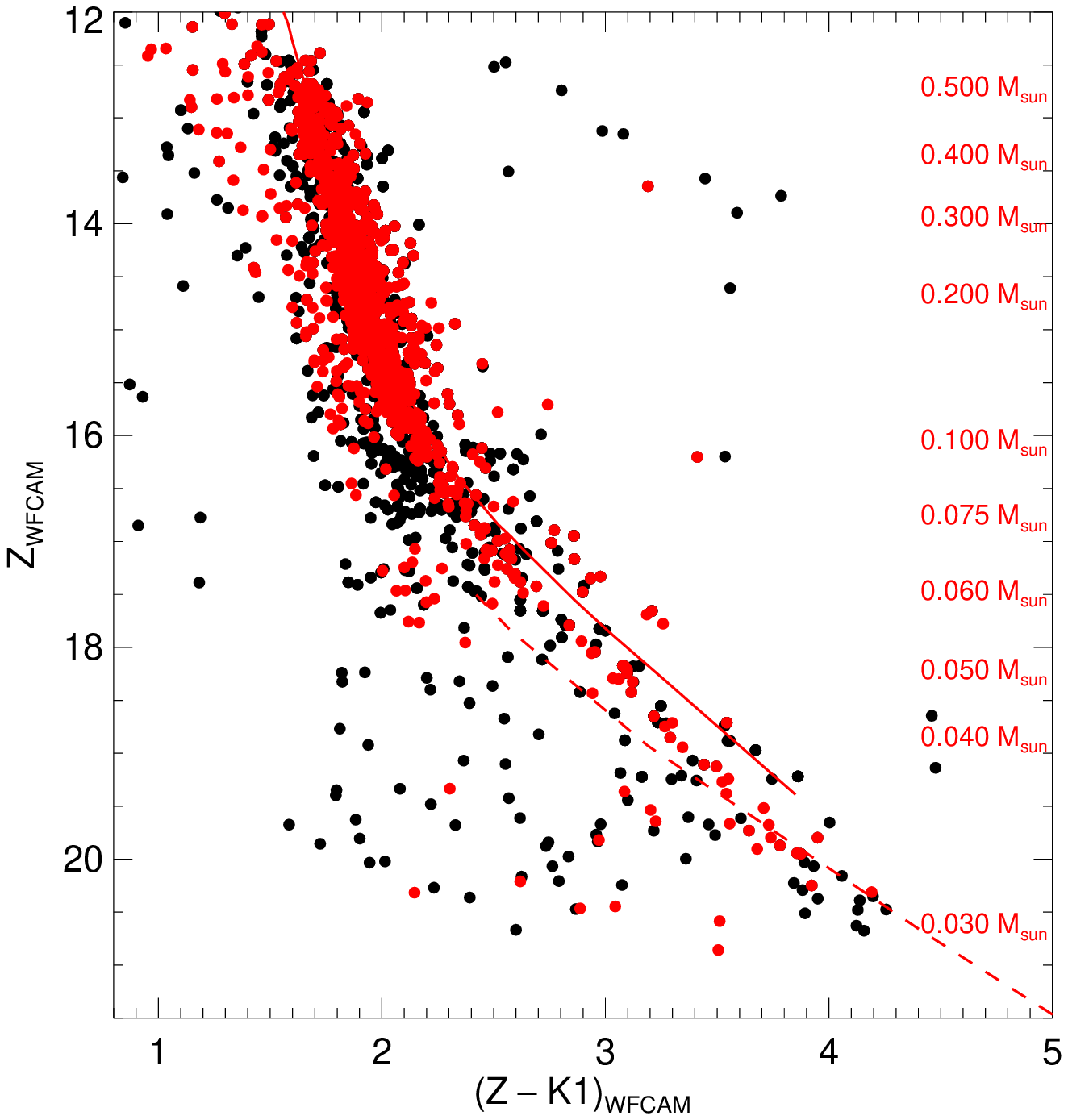}
   \includegraphics[width=0.49\linewidth]{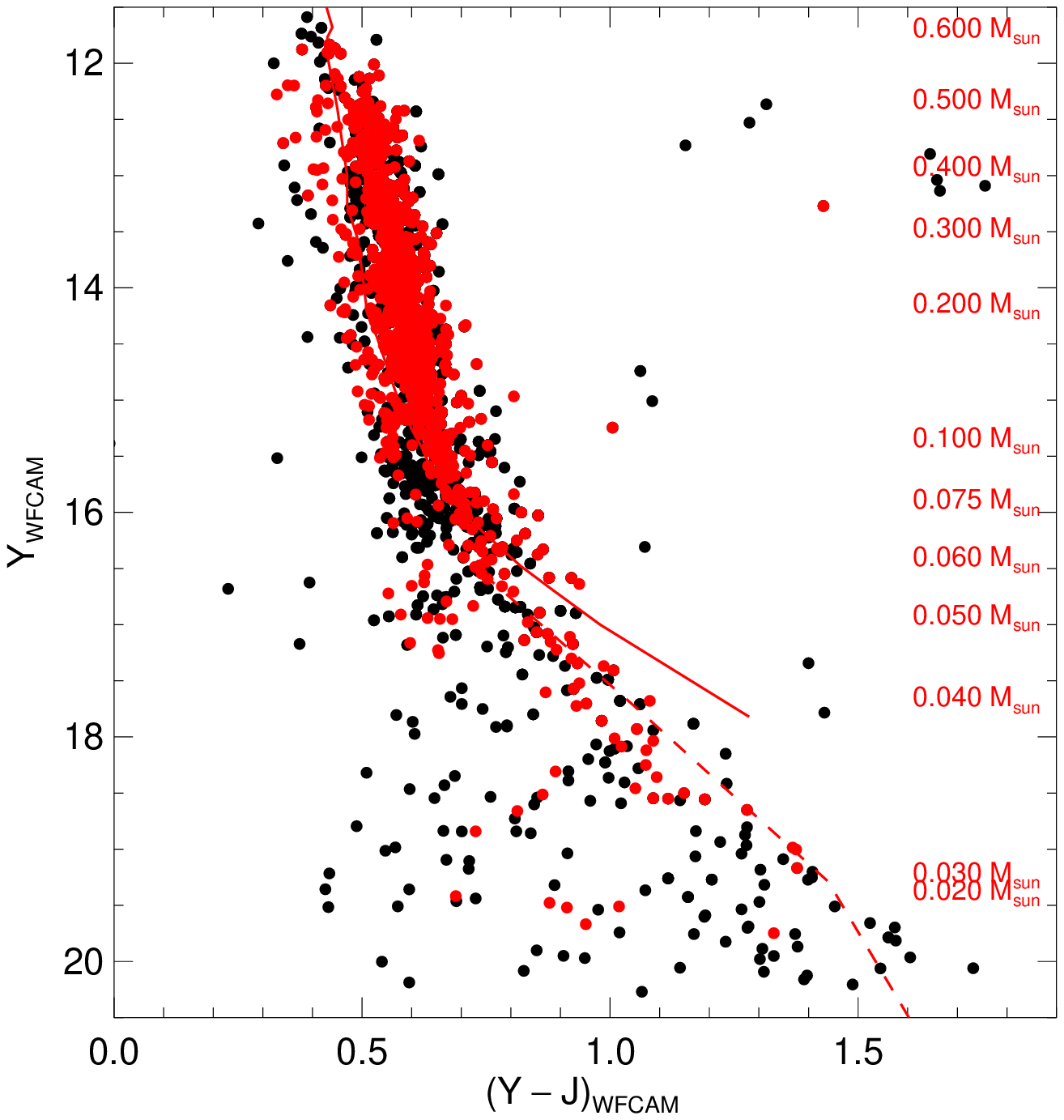}
   \includegraphics[width=0.49\linewidth]{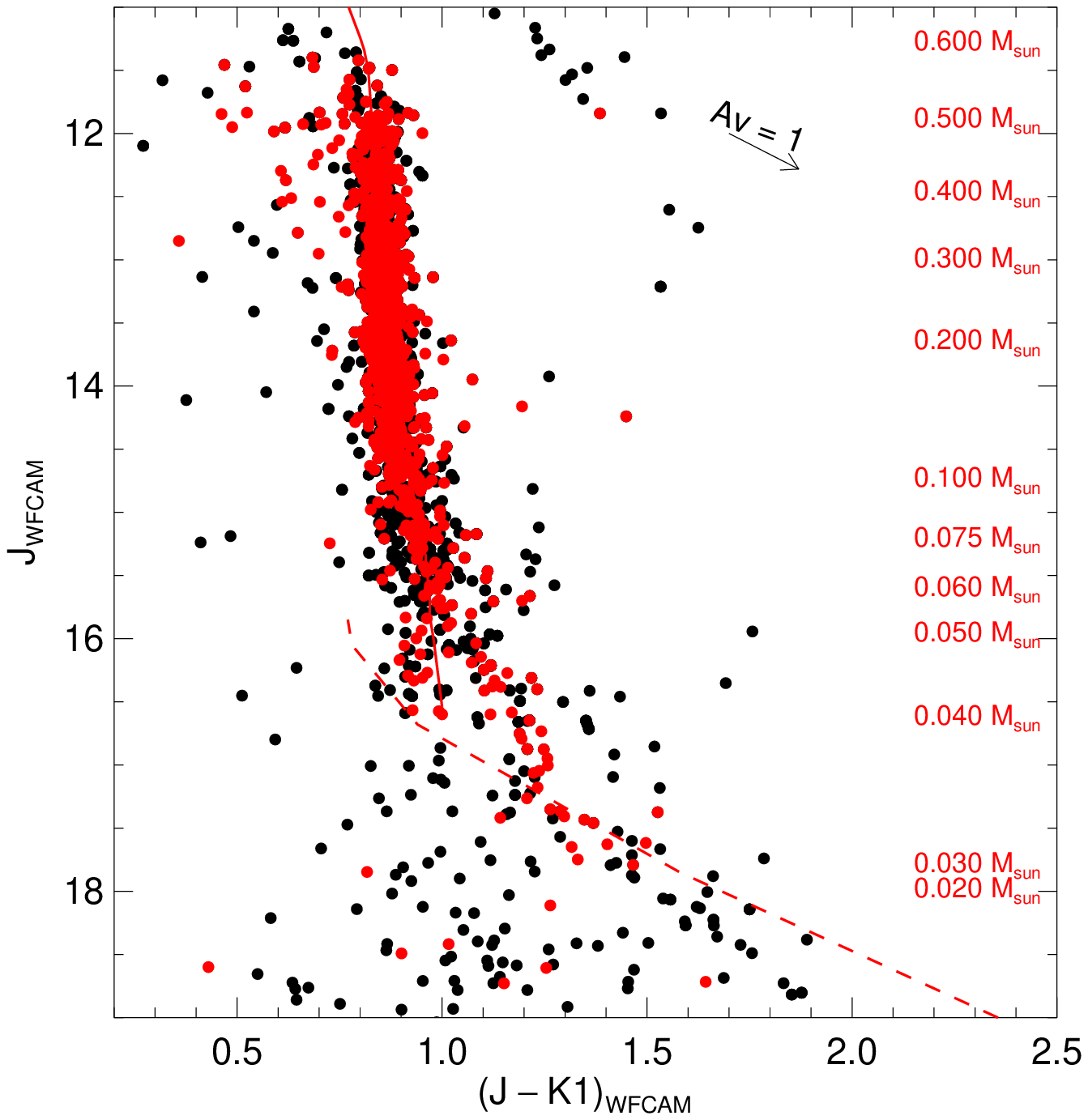}
   \caption{Colour-magnitude diagrams showing the Pleiades member candidates
   previously reported in the literature (black dots) and the new ones extracted
   from our probabilistic analysis (red dots).
   {\it{Upper left:}} ($Z-J$,$Z$);
   {\it{Upper right:}} ($Z-K$,$Z$);
   {\it{Lower left:}} ($Y-J$,$Y$);
   {\it{Lower right:}} ($J-K$,$J$).
   Overplotted are the 120 Myr NextGen \citep[solid line;][]{baraffe98} and
   DUSTY \citep[dashed line;][]{chabrier00c} isochrones shifted to a distance
   of 120 pc. The mass scale is shown on the right hand side of the diagrams
   and spans $\sim$0.6--0.02 M$_{\odot}$, according to the 120 Myr isochrone models.
}
   \label{fig_Pleiades:YJK_cmds}
\end{figure*}

The probabilistic approach yielded a total sample of 8797 sources with
membership probabilities assigned to each of them. This sample contains
947 sources with membership probabilities higher than 60\% listed in
Table \ref{tab_Pleiades:new_members}. Relaxing the probability to 50\%
yields a sample of 1076 Pleiades member candidates.
These high-probability members are displayed in Fig.\ \ref{fig_Pleiades:YJK_cmds}
along with previously published Pleiades candidates.

We note that the sequence of member candidates in our probabilistic
sample is very similar to the sequence of previously-known members down
to $Z$ = 16 mag (Fig.\ \ref{fig_Pleiades:YJK_cmds}). However, we observe
differences in the $Z$ = 16--17 mag range and at fainter magnitudes, mainly
because candidates identified in earlier surveys focusing on Pleiades brown
dwarfs did not have as much as information as the GCS i.e.\ 6-band photometry
and accurate astrometry down to $\sim$25 Jupiter masses. Hence, our new dataset
allows us to reject many of the earlier substellar candidates either on
photometric or astrometric grounds.

%
%

%
%
\begin{figure*}
   \centering
   \includegraphics[width=0.49\linewidth]{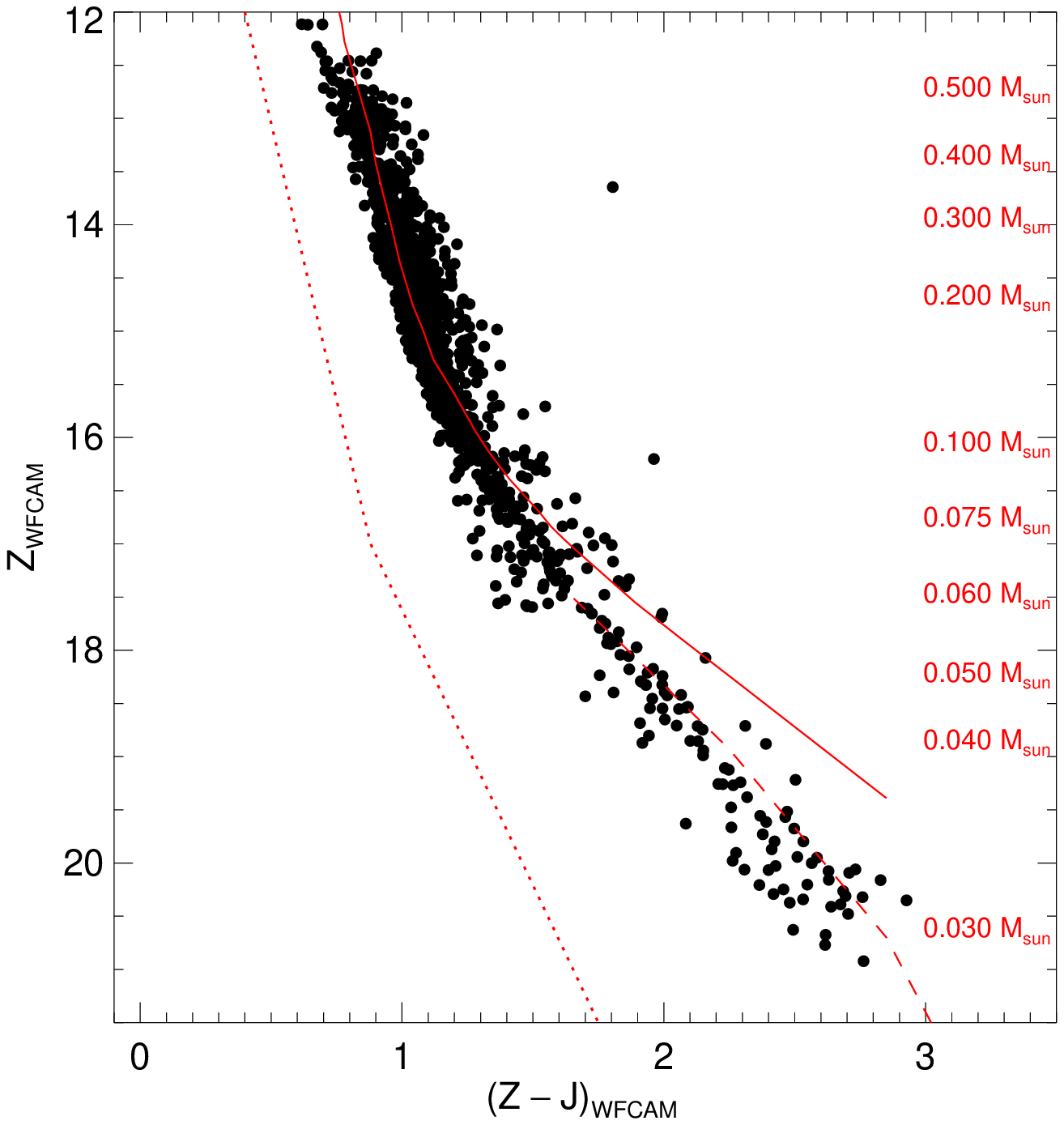}
   \includegraphics[width=0.49\linewidth]{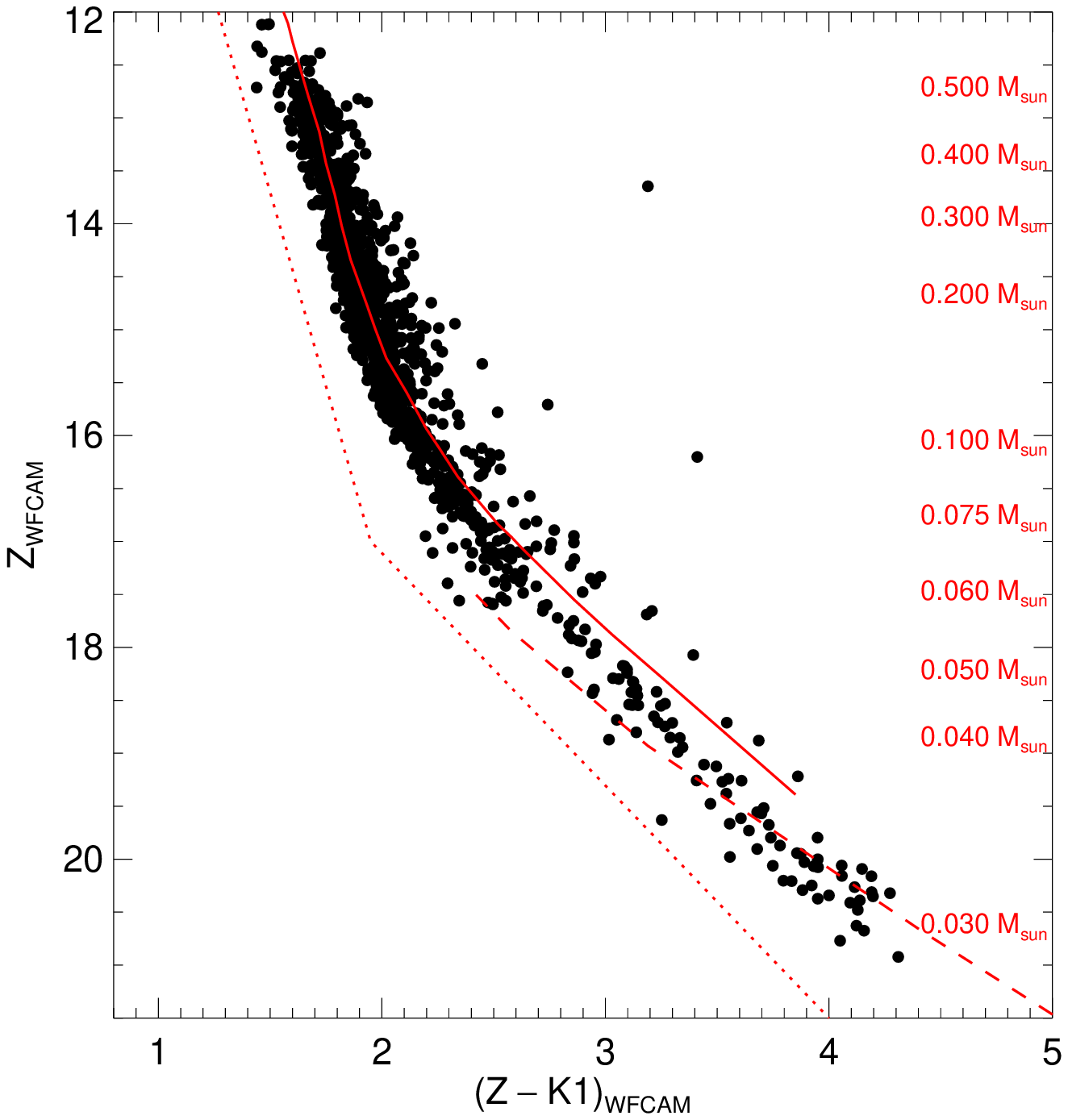}
   \includegraphics[width=0.49\linewidth]{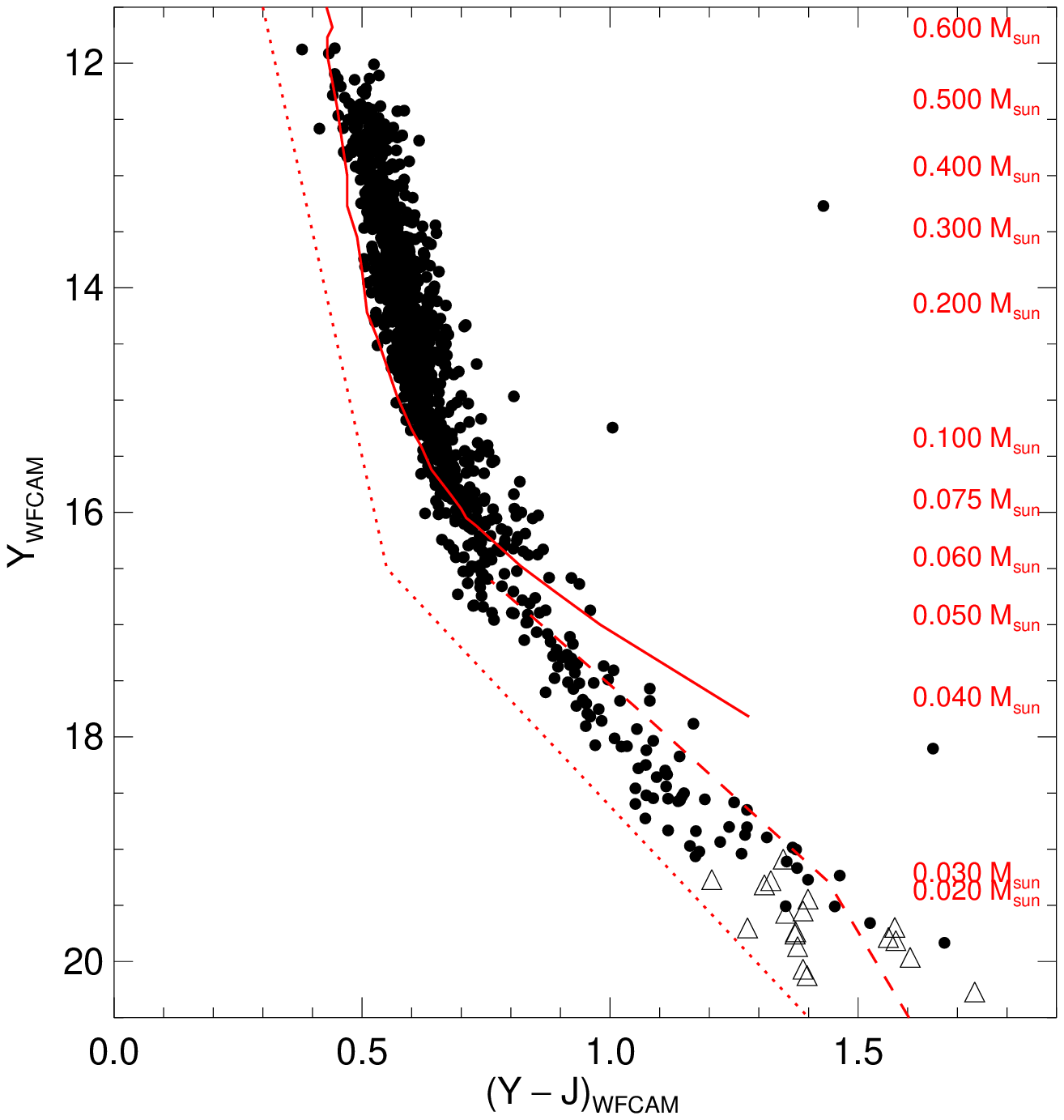}
   \includegraphics[width=0.49\linewidth]{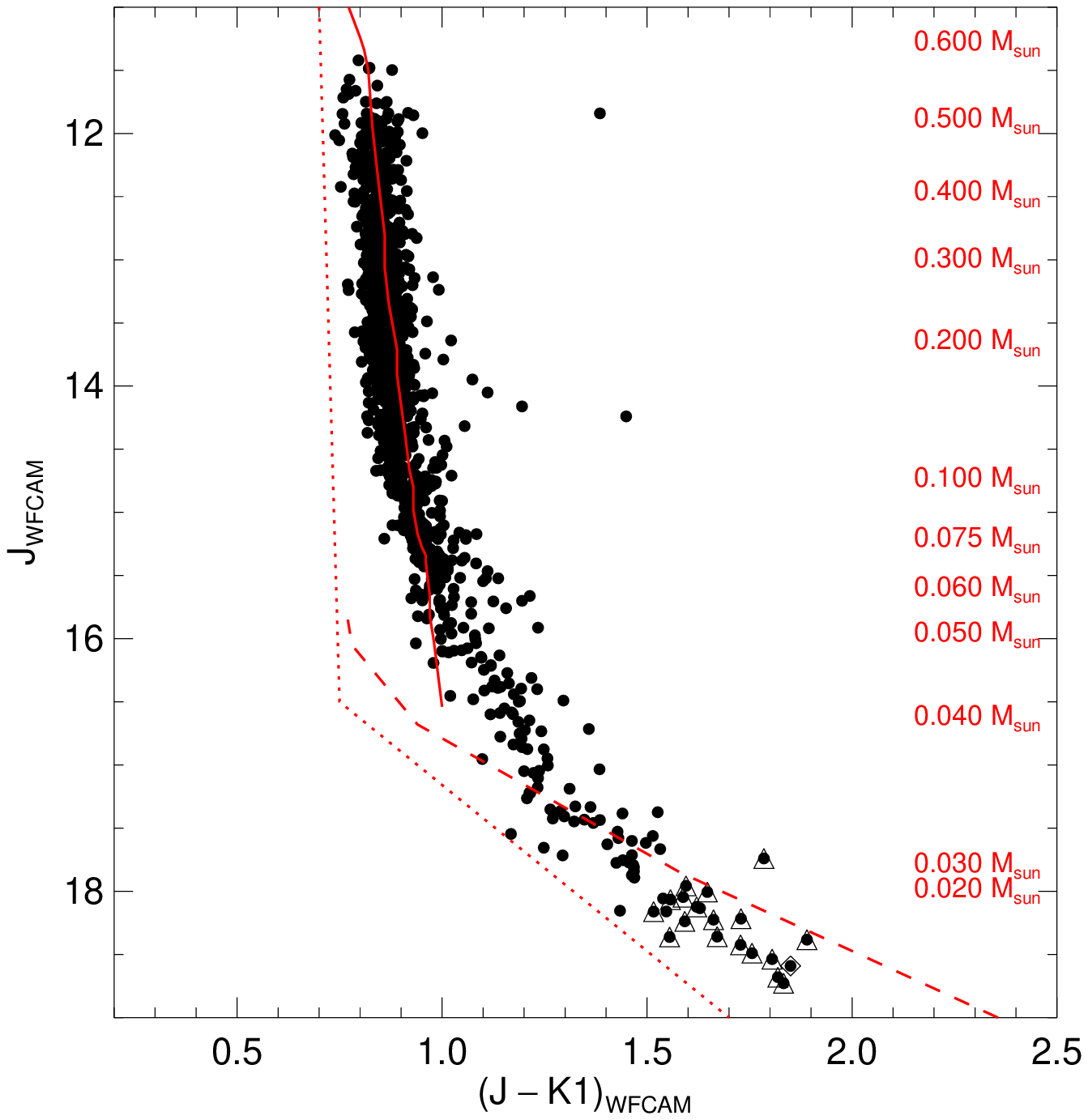}
   \caption{Same as Fig.\ \ref{fig_Pleiades:YJK_cmds} but only for Pleiades 
member candidates selected using method \#2\@. The $YJHK$ and $JHK$-only
detections have been added too. Our photometric criteria listed in 
Section \ref{Pleiades:new_cand_phot_PM} are represented by dotted lines.
The faint $YJHK$ and $JHK$ detections are highlighted with triangles and
diamonds surrounding the black dots, respectively.
}
   \label{fig_Pleiades:YJK_cmds_method2}
\end{figure*}
\subsection{Photometry and proper motion selection}
\label{Pleiades:new_cand_phot_PM}

In this section we outline a more widely used method (referred to as method \#2 in 
the rest of the paper) that we applied to select low-mass and substellar Pleiades 
member candidates. This procedure consists of selecting cluster candidates by 
applying stricter photometric cuts in various colour-magnitude diagrams supplemented 
by a proper motion selection \citep[e.g.][]{lodieu06,lodieu07a}. 
This method is complementary but independent from the probabilistic approach 
presented in the previous section.

First, we plotted several colour-magnitude diagrams 
(Fig.\ \ref{fig_Pleiades:YJK_cmds}) to study the position of known Pleiades
members identified in earlier studies and published over the past decades
(Table \ref{tab_Pleiades:early_summary}). Based on these known members,
we define a series of lines to select photometrically member candidates with
photometry in $ZYJHK$ in four colour-magnitude diagrams as indicated below
(dotted lines in the diagrams in Fig.\ \ref{fig_Pleiades:YJK_cmds_method2}):
\begin{itemize}
\item ($Z-J$,$Z$) = (0.50,12.0) to (1.05,17.0) 
\item ($Z-J$,$Z$) = (1.05,17.0) to (2.40,21.5) 
\item ($Z-K$,$Z$) = (1.20,11.5) to (1.95,17.0)
\item ($Z-K$,$Z$) = (1.95,17.0) to (4.00,21.5)
\item ($Y-J$,$Y$) = (0.30,11.5) to (0.55,16.5) 
\item ($Y-J$,$Y$) = (0.55,16.0) to (1.40,20.5) 
\item ($J-K$,$J$) = (0.70,11.0) to (0.70,16.5)
\item ($J-K$,$J$) = (0.70,16.5) to (1.70,19.0) 
\end{itemize}

These photometric cuts remain conservative and the contamination to the blue
side of the Pleiades sequence is still high. Hence, the second step consisted of
applying a proper motion selection in the vector point diagram 
(Fig.\ \ref{fig_Pleiades:diagram_VPD}) to improve on the photometric selection.
We applied a 3$\sigma$ selection given the formal errors on the individual proper 
motions for each object, implying a completeness higher than 99\% for normally--distributed errors.
We assumed a mean proper motion of ($+$18, $-$42) mas/yr for the Pleiades,
slightly different from the Hipparcos values \citep{robichon99,vanLeeuwen09}
because the proper motion measurements are on a relative system rather
than the Hipparcos system as described in Collins \& Hambly (2012).
The main advantage of this method is that it doesn't rely on a single radius
for the proper motion selection but rather takes into account the increasing
uncertainty on the proper motion measurements for fainter stars, and allows for
different time baselines of the epoch frames affecting the proper motion errors.

After applying both the photometric and proper motion selections, we found a
large number of objects lying to the blue of the Pleiades sequence. Hence, we
applied an additional photometric cut in the ($Z-J$,$Z$) colour-magnitude
diagram, eliminating all sources in the $Z$ = 12--18 mag range and located
to the left of a line defined by ($Z-J$,$Z$) = (0.6,12.0) to (1.2,16.5).
This selection yielded a total 1147 low-mass stars and brown dwarfs with
$Z$ magnitude ranging from 12 to 21.5 (Table \ref{tab_Pleiades:new_members}).
This total number is similar to the number of high probability Pleiades
member candidates identified via the probabilistic approach; we note however that
the membership count from the different methods is identical within the counting 
errors if we sum the membership probabilities of {\em all} stars, as expected.

%
%
\subsection{Search for lower mass members}
\label{new_cand_lower_mass}

In this section we search for fainter and cool substellar members of the Pleiades
by dropping the constraint on the $Z$-band detection and later the $Z+Y$ bands.

\subsubsection{$YJHK$ detections}

To extend the Pleiades cluster sequence to fainter brown dwarfs and cooler
temperatures, we searched for potential candidate members undetected in $Z$.
We imposed similar photometric and astrometric criteria as those detailed in 
Section \ref{Pleiades:new_cand_phot_PM} but imposed a $Z$ non detection and
associated criteria as described below:
\begin{itemize}
\item No $Z$ detection
\item $Y$ $\geq$ 18 and $J$ $\leq$ 19.3 mag
\item Candidates should lie above the line defined by ($Y-J$,$Y$) = (0.55,16.0) 
and (1.40,20.5)
\item Candidates should lie above the line defined by ($J-K$,$J$) = (0.75,16.5) 
and (1.70,19.0)
\item The position on the proper motion vector point diagram of each candidate 
should not deviate from the assumed cluster proper motion by more than 3$\sigma$
\end{itemize}
This selection returned 22 additional Pleiades member candidates
which have been added to Table \ref{tab_Pleiades:new_members}
along with sources identified with both selection methods presented earlier.
All but four of them are indeed invisible in the $Z$-band images and look well 
detected in the other bands after checking the GCS DR9 images by eye.

\subsubsection{$JHK$ detections}

We repeated the procedure described above looking for $Z$ and $Y$ non detections.
We applied the following criteria:
\begin{itemize}
\item No $Z$ and $Y$ detection
\item $J$ = 18--19.3 mag
\item Candidates should lie above the line defined by ($J-K$,$J$) = (0.75,16.5)
and (1.70,19.0)
\item The position on the proper motion vector point diagram of each candidate 
should not deviate from the assumed cluster proper motion by more than 3$\sigma$
\end{itemize}

This query returned 19 new Pleiades candidate members. After checking the images by eye, 
we kept only one of them because one is actually visible in $Y$ (although a $Y$
detection is not reported in the GCS DR9 catalogue), one is visible in both
the $Z$ and $Y$ images. The remaining 16  
have no $Z$ and $Y$ images available so we can't confirm if they are indeed
drop-outs (Table \ref{tab_Pleiades:new_members}).

%
%
\section{The substellar binary frequency}
\label{Pleiades:binary}

The multiplicity in the substellar domain at different ages provides one way to 
constrain the formation mechanisms of brown dwarfs. As in our earlier study
of the Pleiades \citep{lodieu07c}, we investigated the binary frequency of Pleiades 
brown dwarfs using the photometry and colours from the GCS\@. However, our sample
is now two times larger and, with the proper motions, is of much higher quality 
because it is drawn from the same homogeous survey. We consider in
this section all Pleiades member candidates selected through method \#2\@.

\begin{figure*}
   \includegraphics[width=0.49\linewidth]{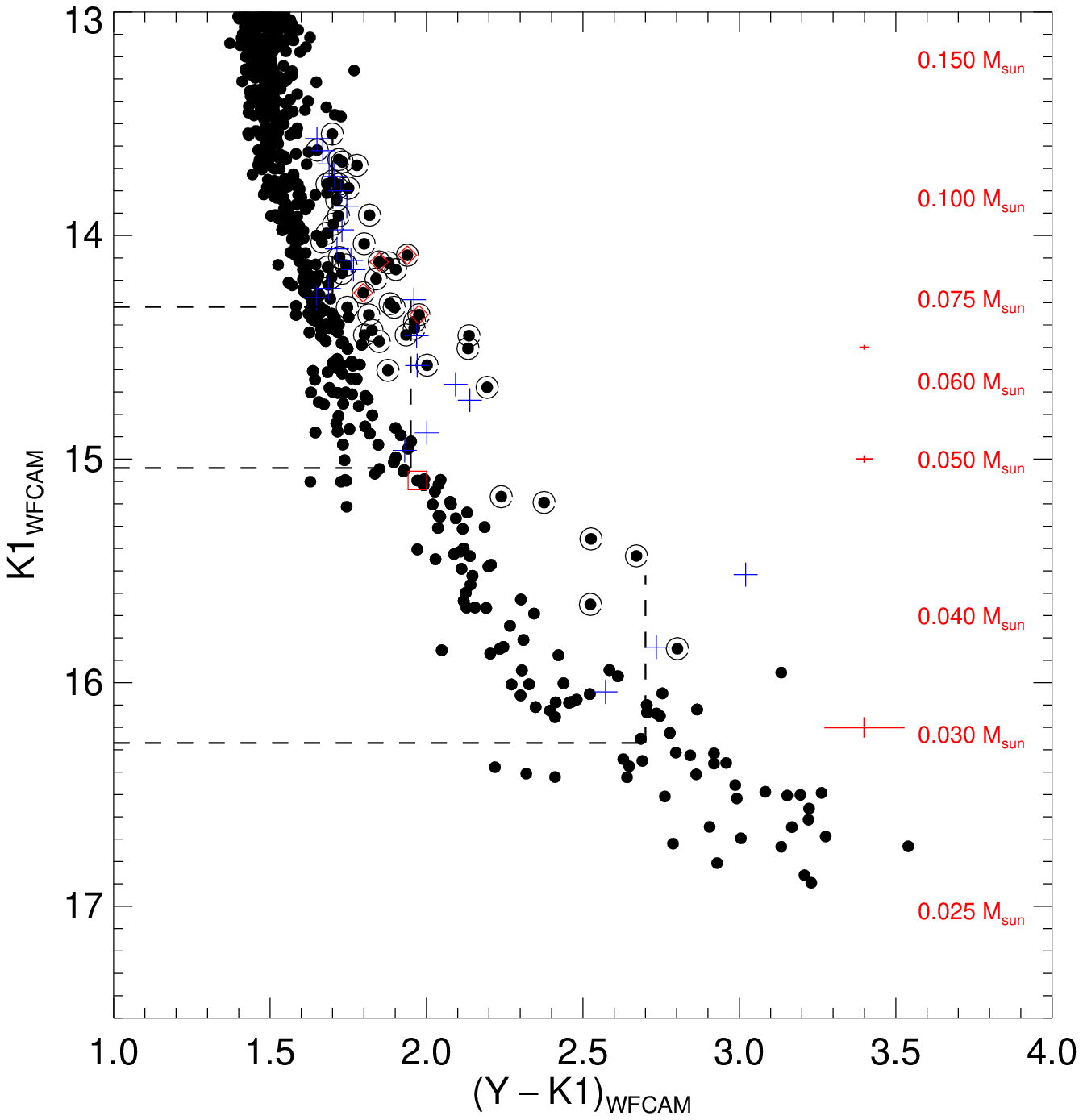}
   \includegraphics[width=0.49\linewidth]{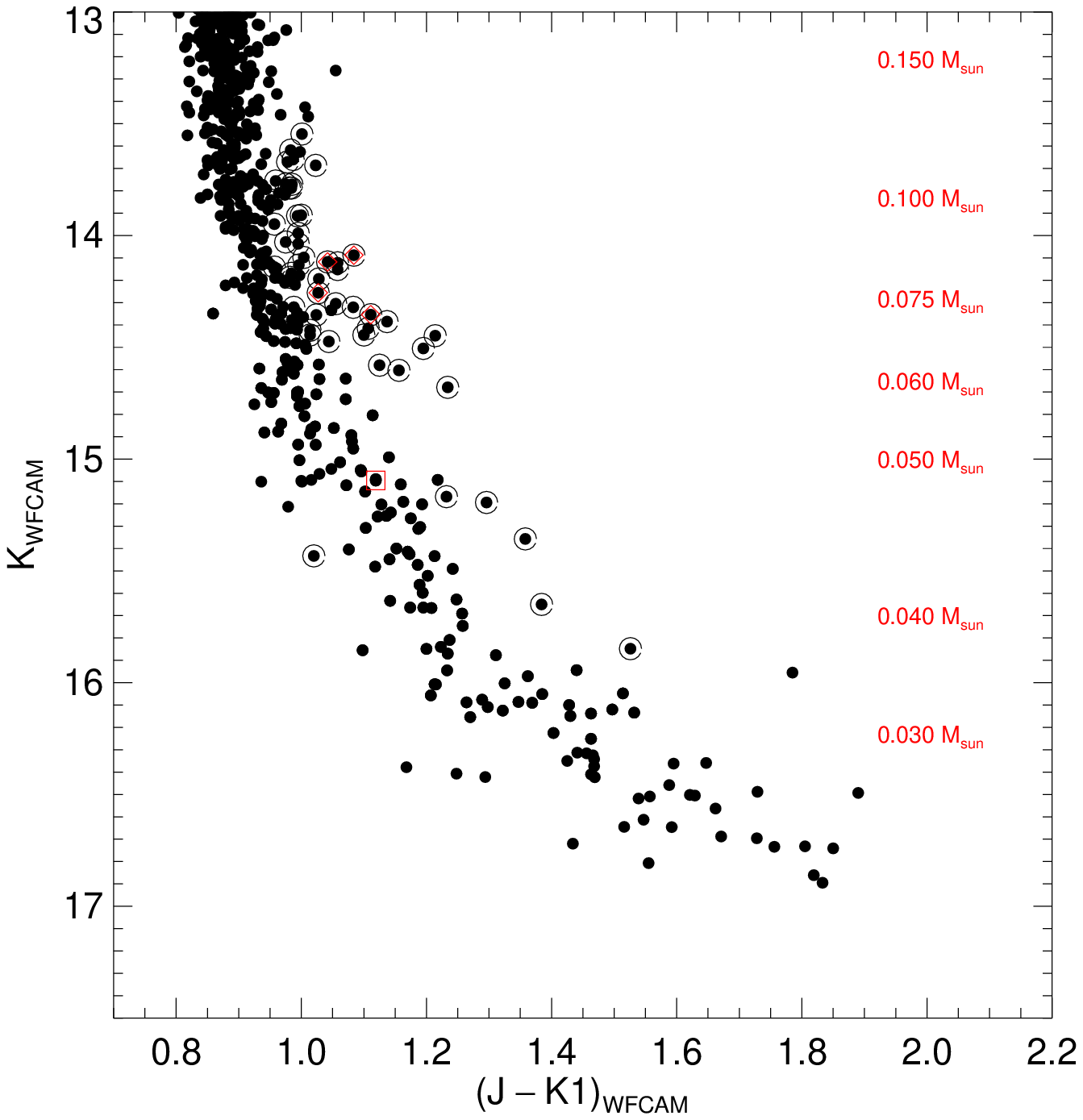}
   \caption{($Y-K$,$K$) and ($J-K$,$K$) colour-magnitude diagrams showing all
Pleiades member candidates selected with method \#2 as well as the $YJHK$
and $JHK$-only detections. The multiple system candidates identified 
photometrically in the ($Y-K$,$K$) diagram are marked with a circle; the red square
shows the location of the prototype Pleiades brown dwarf Teide 1\@.
The mass scale is shown on the right hand side of the diagrams for an
age of 120 Myr and a distance of 120.2 pc. Photometric error bars are shown
as crosses on the right-hand side of the panel. The method describing the
selection of binary candidates in the ($Y-K$,$K$) diagram is highlighted
with black dashed lines. Blue crosses represent the expected sequences of
binary systems for primaries with masses of 0.075, 0.05, and 0.03 M$_{\odot}$,
respectively.
} 
   \label{fig_Pleiades:diagram_BF}
\end{figure*}

We applied the same method as described in \citet{lodieu07c} to select substellar
binary candidates in the Pleiades. We briefly summarise the technique here.
Figure \ref{fig_Pleiades:diagram_BF} displays two colour-magnitude diagrams
used to identify binary candidates because of the large colour range and the
presence of sources above the cluster (single star) sequence. We started off
our selection in the ($Y-K$1,$K$1) colour-magnitude diagram (left-hand side plot
in Fig.\ \ref{fig_Pleiades:diagram_BF}) because it shows large colour difference
in the substellar regime. We applied the following method:
for a given magnitude, e.g.\ $K$1 = 15.5--16.5 mag, we defined two horizontal 
lines intercepting the mean value of the single object sequence. From the 
intercept points we defined two vertical lines with a length of 0.75 mag
(dashed lines in Fig.\ \ref{fig_Pleiades:diagram_BF}).
Then we divided the box formed by both sequences and the vertical lines 
into two boxes: single stars lie in the bottom part whereas binary candidates 
in the top one. Except for one system which appears to the blue of the single-star
sequence in the ($J-K$1,$K$1) diagram, the location of the binary candidates is 
confirmed in this diagram and others as well, adding credence to their potential 
multiplicity. Our method is corroborated by the presence of four known Pleiades 
brown dwarf binaries \citep[IPMBD\,25, IPMBD\,29, PPl15, CFHT-Pl-IZ\,4][]{basri99b,martin00a,martin03,bouy06a}
above the single star sequence where Teide 1 \citep{rebolo95}, an isolated Pleiades 
brown dwarf is located (red symbol in Fig.\ \ref{fig_Pleiades:diagram_BF}).
Note that we observe a dispersion of 0.5~mag in the single-star sequence which 
can explained by the tidal radius of the cluster leading to a variation of 10\% 
in the distance of the members, i.e.~$\sim$12~pc corresponding to $\sim$0.2~mag
\citep{pinfield00}.

The binary fraction was then defined as the number of binaries divided by the 
total number of objects (single stars$+$binaries). We counted 51 binary candidates
(Table \ref{tab_Pleiades:binary_candidates}) and 137 single stars in the 
$K$1 $\sim$ 14.32--16.27 mag range, corresponding to masses between 0.075 and 
0.03 M$_{\odot}$ at the age and distance of the Pleiades. Hence, we derive a 
binary frequency of 51/(137$+$51) = 27.1$\pm$5.8\% in this mass range for projected 
separations smaller than $\sim$100--200 au. This value is likely a lower limit 
because some binaries may hide in the single star sequence due to higher 
separations or mass ratios and we have not considered wider systems. The overall
result is in agreement within the error bars with our previous estimate although 
on the lower side \citep[36.5$\pm$8.0\%][]{lodieu07c}. We also 
divided up this mass range into two bins covering 0.075--0.05 M$_{\odot}$
and 0.05--0.03 M$_{\odot}$, yielding binary fractions of 33.0$\pm$9.1\% 
(35 binaries and 71 singles) and 19.5$\pm$7.0\% (16 binaries and 66 singles), 
respectively. The binary frequency over the lowest mass range is not reproduced 
by the latest hydrodynamical
simulation of a 500 M$_{\odot}$ stellar cluster as no brown dwarf binary
was found in the 0.07--0.03 M$_{\odot}$ range \citep[Table 2 in][]{bate11b}.
Unfortunately, we cannot test further those theoretical predictions with
estimates on the numbers of triple and high-order multiple, separation
distributions, and mass ratios.

We investigated the range of validity of our binary frequency by plotting 
the expected positions of binary systems with primary masses of 0.075,
0.05, and 0.03 M$_{\odot}$ (blue crosses in Fig.\ \ref{fig_Pleiades:diagram_BF}.
Adding smaller mass brown dwarfs (going from 0.075 M$_{\odot}$ i.e.\ equal-mass
binaries down to 0.02 M$_{\odot}$) to primaries with masses of 0.075 M$_{\odot}$
places those systems to the red of single stars and then turn over towards
higher luminosities. The same behaviour is observed for binaries with
primary masses of 0.05 M$_{\odot}$ whereas binaries with primary masses
of 0.03 M$_{\odot}$ turn redder and brighter. To estimate the sensitivity
of our binary frequency in terms of mass ratios, we considered the photometric
errors of the GCS for the three mass values. We inferred that our binary
frequencies over the 0.075--0.05 M$_{\odot}$ and 0.05--0.03 M$_{\odot}$ are 
valid for mass ratios larger than 0.4--0.5 and 0.8, respectively. Hence,
the factor of two observed in the multiplicity between those two mass bins
should be interpreted with caution because they are not valid over the same
mass ratio interval. If we consider only binaries with mass ratios larger
than 0.8 in the 0.075--0.05 M$_{\odot}$ mass range, we derive a binary
fraction of 9/(71$+$9) = 11.3$\pm$5.0\%, lower by a factor of two than the
frequency inferred in the 0.05--0.03 M$_{\odot}$ range although consistent
within the error bars.

We note that we applied the same procedure to the probabilistic sample
and found very similar numbers which do not change the main conclusions discussed
in this Section. We infer a substellar binary frequency of 24.3$\pm$7.3\%
which can be divided up into 29.2$\pm$12.0 and 20.0$\pm$8.7\% for the
0.075--0.05 M$_{\odot}$ and 0.05--0.03 M$_{\odot}$ mass bins, respectively.

Our binary fraction is higher by a factor two to three than the frequency inferred 
from high-resolution imaging 
\citep[13.3$^{+13}_{-4}$\%;][]{martin00a,martin03,bouy06a} especially if
we consider that all known Pleiades brown dwarf binaries lie in the
0.065--0.055 M$_{\odot}$ mass range. Our multiplicity is lower than the 50\% 
estimate derived by \citet{pinfield00} from a purely photometric estimate (i.e.~no
proper motion measurements involved) and on the low side of Monte-Carlo
predictions \citep[32--45\%;][]{maxted05}. However, our results are in agreement 
with the upper limit of 26$\pm$10\%  quoted by \citet{basri06} for low-mass stars 
and brown dwarfs. This total binary frequency is split into 11\% of spectroscopic
binaries with projected separation below 3 au, and 15\% of wider binaries 
(3--15 au) in agreement with high-resolution imaging surveys of field ultracool
dwarfs \citep[for a review, see][]{burgasser07a} and theoretical predictions
from hydrodynamical simulations \citep{bate11b}.

%
%
%

%
%
%
\begin{figure}
   \includegraphics[width=\linewidth]{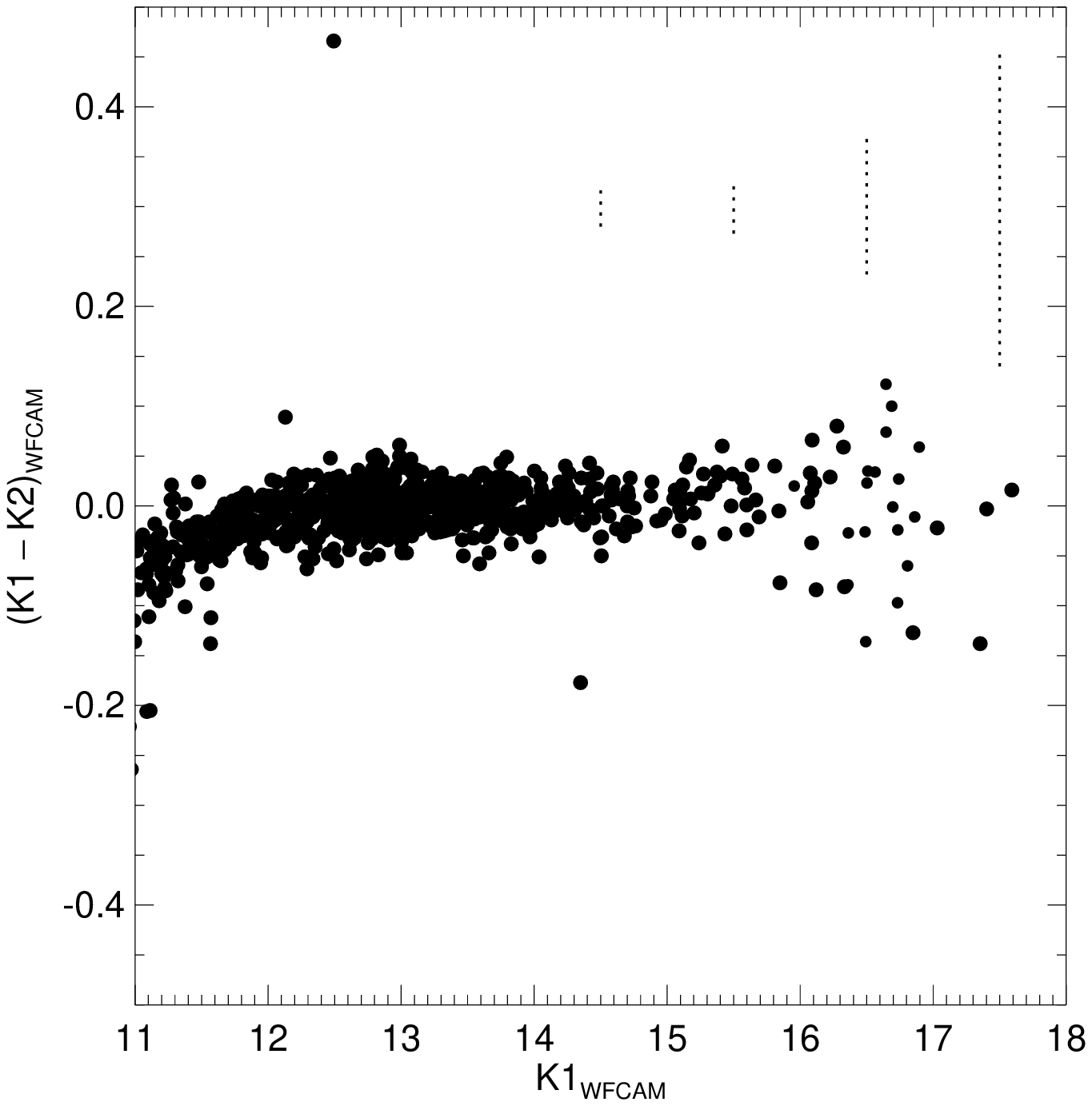}
   \caption{Difference in the $K$ magnitude ($K$1--$K$2) as a function of the 
$K$1 magnitude for all Pleiades member candidates with membership probability 
greater than 60\%. The $YJHK$ and $JHK$-only detections have been added too.
Typical error bars on the colour shown as vertical dotted lines are added 
at the top of the plot.
} 
   \label{fig_Pleiades:diagram_variability}
\end{figure}
%

%
%
\section{Variability at 120 Myr}
\label{Pleiades:variability}

In this section we discuss the variability of the Pleiades low-mass stars
and brown dwarfs using the two epochs provided by the GCS\@. We only considered 
the Pleiades member candidates with membership probabilities larger than 60\%,
many of them being already published in the literature
(Tables \ref{tab_Pleiades:early_DR9} and \ref{tab_Pleiades:new_members}).

Figure \ref{fig_Pleiades:diagram_variability} shows the ($K$1--$K$2) vs $K$1 
for all Pleiades member candidates with probabilities higher than 60\%. The
brightening in the $K$1 = 11--12 mag range is due to the difference in saturation characteristics
between the first and second epoch, of the order of 0.5 mag both in the
saturation and completeness limit. This is understandable because the exposure
times have been doubled for the second epoch with relaxed constraints on the
seeing requirement and weather conditions. We excluded those objects from
our variability study. Overall, the sequence indicates consistent photometry
between the two K epochs and very few objects appear variable in the $K$-band.

We selected variable objects by looking at the standard deviation,
defined as 1.48$\times$ the median absolute deviation which is the median of 
the sorted set of absolute values of deviation from the central value of
the $K$1--$K$2 colour. In the $K$1 = 11--13 mag, we identified four potential
variable abjects with differences in the $K$-band larger than 3$\sigma$ above
the standard deviation. However, all appeared saturated in the second epoch
images suggesting that the variability may be caused by the inaccurate photometry
derived from saturated sources. We extracted another potential variable
low-mass star around $K$\,$\sim$14.5 mag but this source is located at the edge
of the detector, casting doubt on any intrinsic variability.
The analysis is not possible beyond 16 mag due to the small number of Pleiades
members with high probabilities.

We conclude that the level of K--band variability at 120 Myr is small, with standard
deviations in the 0.05--0.08 mag range, suggesting that it cannot account for 
the dispersion in the cluster sequence. The same conclusions are drawn from
the sample derived from method \#2\@. 

%
%
\section{The Initial mass function}
\label{Pleiades:IMF}

In this section we discuss the cluster luminosity and mass functions derived from 
the samples of Pleiades member candidates extracted from both methods described
in the previous section. We did not attempt to correct the mass function for 
binaries.

\subsection{The cluster luminosity function}
\label{Pleiades:IMF_LF}

In this section, we construct two luminosity functions: i) we used the sample 
of 8797 Pleiades cluster member candidates selected by the probabilistic 
approach (Section \ref{Pleiades:new_cand_probabilistic}); and ii), the 1147 
candidates identified with method \#2 (Section \ref{Pleiades:new_cand_phot_PM}).
The luminosity function of the former method is derived by summing membership 
probabilities of all stars fitted to distribution functions in the vector 
point diagram, whereas the luminosity function of the latter is derived 
simply by summing the number of member candidates.

\begin{figure*}
   \includegraphics[width=0.49\linewidth]{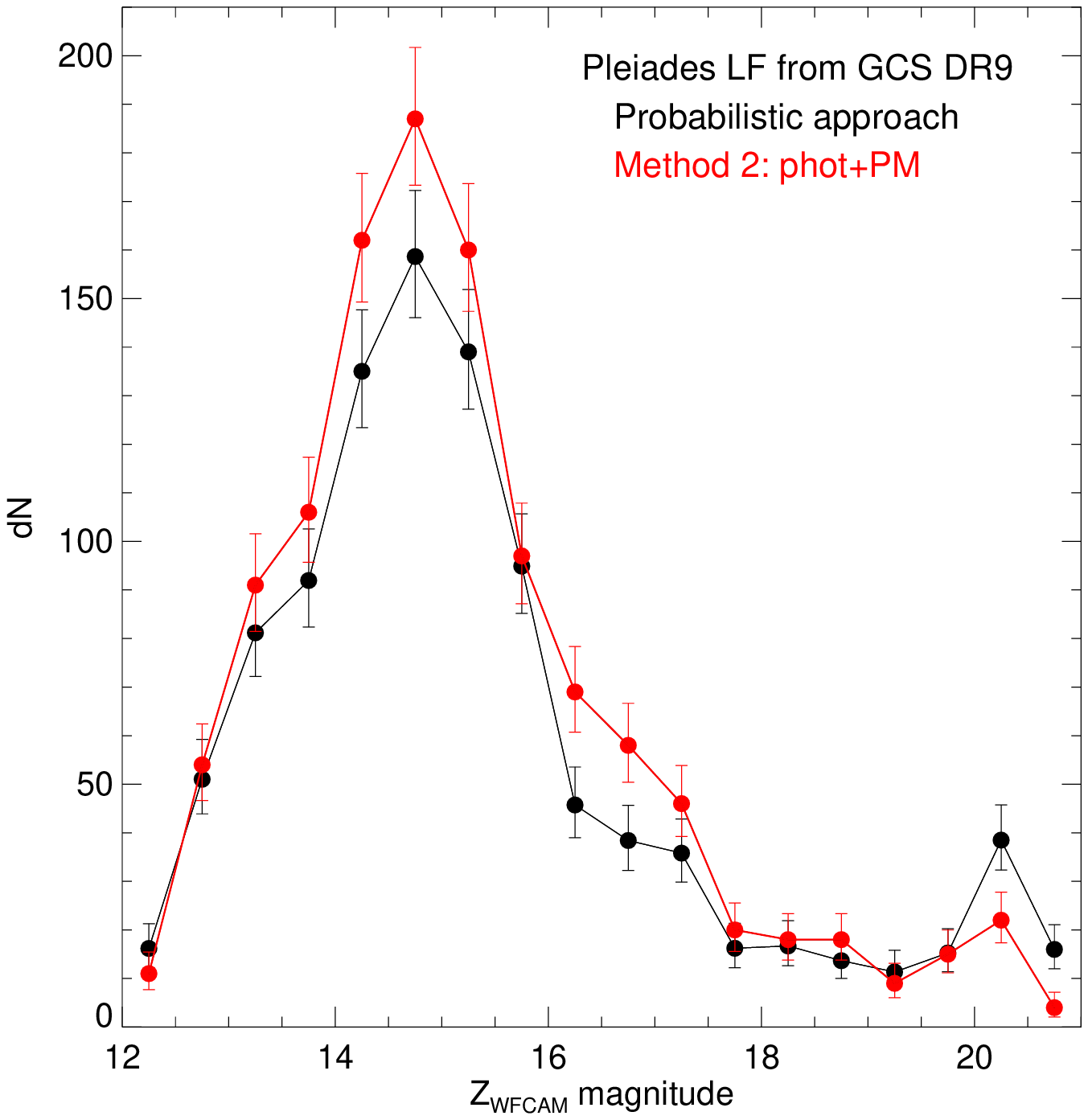}
   \includegraphics[width=0.49\linewidth]{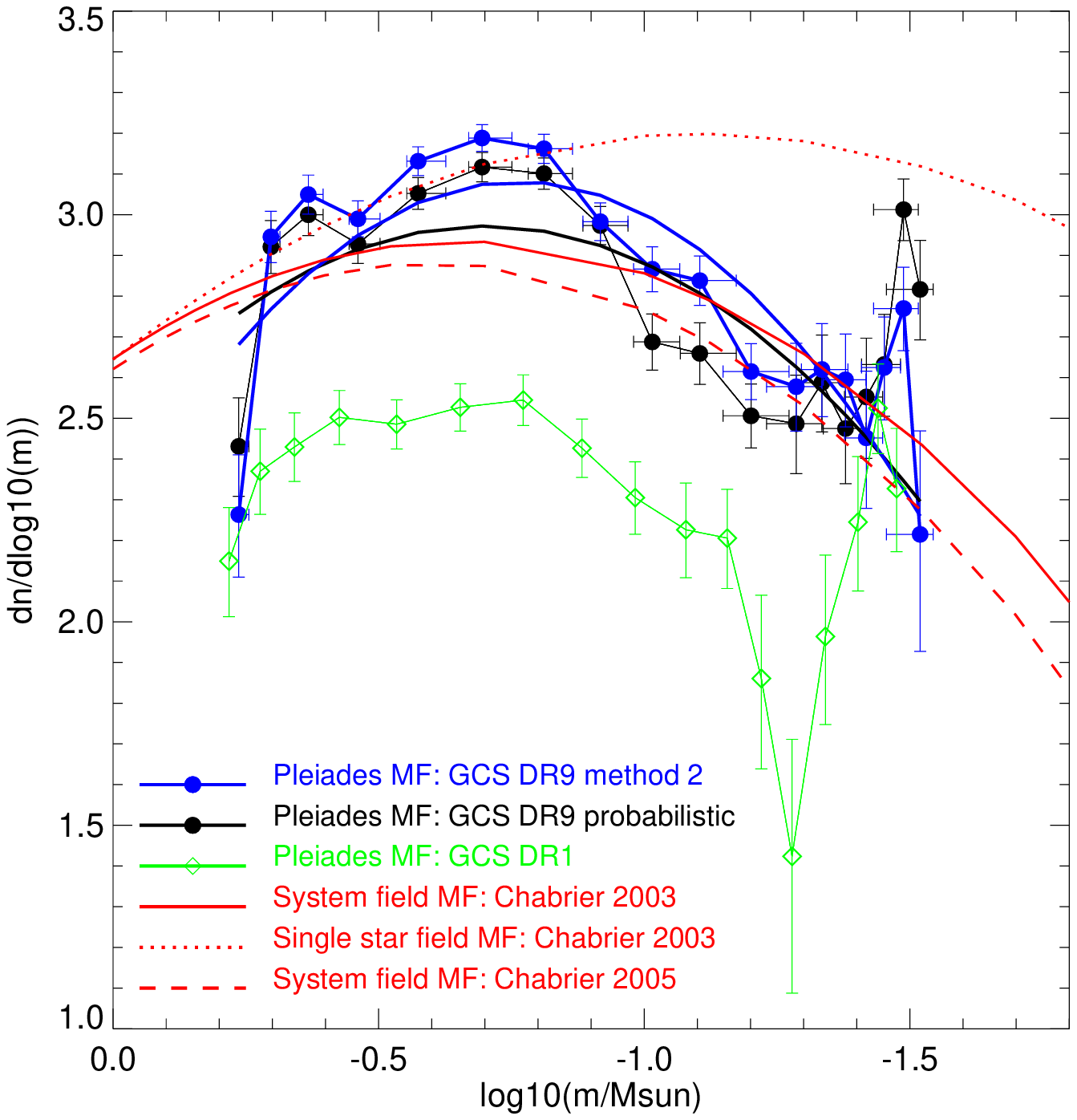}
   \caption{Luminosity (left) and mass (right) functions derived from our 
analysis of the UKIDSS GCS DR9 sample of Pleiades member candidates.
The left-hand side panel compares the luminosity function obtained from the 
probabilistic approach (black symbols and black line) and the luminosity 
function derived from the selection outlined by method \#2 (red colour).
Error bars on the number of objects per magnitude bin are Gehrels errors.
The right-hand side panel compares the Pleiades mass function derived from the
probabilistic approach (filled black dots linked by a solid line) and the mass 
function derived from method \#2 (blue symbols and blue line). The black and 
blue lines are log-normal fits to the Pleiades mass function. Error bars on the 
dN/d$\log$M are Gehrels errors whereas error bars on the masses come from the 
3$\sigma$ lower and upper limits on the age (100--150 Myr) and distance 
(114.5--125.9 pc). The first and last two points of the Pleiades luminosity
and mass functions are to be treated with caution due to saturation and 
contamination at the bright and faint ends, respectively.
The mass function from our GCS DR1 analysis (green diamond and green line) and 
the field mass functions \citep[red lines;][]{chabrier03,chabrier05a} are also 
included for completeness and comparison.
} 
   \label{fig_Pleiades:LF_and_MF}
\end{figure*}

Both luminosity functions i.e.\ the number of stars and brown dwarfs as a 
function of magnitude plotted per 0.5 mag bin is displayed in 
Fig.\ \ref{fig_Pleiades:LF_and_MF}. The brightest bin is a lower limit due 
to the saturation limit of the GCS survey. The last bin is very likely 
incomplete due to the constraint imposed on the $Z$-band detection. The 
numbers of objects per 0.5 mag bin increase quickly to reach a peak around 
$Z$ = 14.5--15 and drop off afterwards down to the completeness of our 
survey with a possible peak beyond $Z$ = 20 mag 
(Tables \ref{tab_Pleiades:LF_MF_probabilistic} \& \ref{tab_Pleiades:LF_MF_method2}). 
Both luminosity functions look very similar and match each other within the 
error bars. Therefore, we conclude that both methods provide the same result 
and a good representation of the Pleiades luminosity and mass functions.

%
%
%

%
%
\begin{table*}
  \caption{Values for the luminosity and mass functions (both in linear and
logarithmic scales) per magnitude and mass bin for the Pleiades open cluster 
using the probabilistic approach (method~1). We assume a distance of 120.2 pc 
and employed the NextGen and DUSTY 120 Myr theoretical isochrones for the 
mean values. The uncertainties given in brackets for the mid-mass (Column 4)
come from the 3$\sigma$ lower and upper limits on the age and distance, assuming
errors of 8 Myr and 1.9 pc, respectively.
}
  \label{tab_Pleiades:LF_MF_probabilistic}
  \begin{tabular}{c c c c | c c c | c c c | c c c}
  \hline
Mag range &   Nb\_obj & Mass range  & Mid-mass & dN   & errH & errL & dN/dM & errH & errL & dN/dlogM & errH & errL \cr
  \hline
12.0--12.5 &  179 & 0.6200--0.5400 & 0.5800 (0.5550--0.5890) &  16.18 &   5.11 &   3.99 &  202.29 &   63.94 &   49.90 &  2.43 & 0.27 & 0.28 \\
12.5--13.0 &  306 & 0.5400--0.4690 & 0.5045 (0.4850--0.5125) &  51.03 &   8.20 &   7.13 &  718.70 &  115.43 &  100.36 &  2.92 & 0.15 & 0.15 \\
13.0--13.5 &  372 & 0.4690--0.3890 & 0.4290 (0.4025--0.4380) &  81.18 &  10.05 &   9.00 & 1014.71 &  125.64 &  112.45 &  3.00 & 0.12 & 0.12 \\
13.5--14.0 &  480 & 0.3890--0.3030 & 0.3460 (0.3145--0.3595) &  91.95 &  10.63 &   9.58 & 1069.15 &  123.58 &  111.35 &  2.93 & 0.11 & 0.11 \\
14.0--14.5 &  614 & 0.3030--0.2300 & 0.2665 (0.2365--0.2800) & 135.02 &  12.65 &  11.61 & 1849.55 &  173.31 &  159.03 &  3.05 & 0.09 & 0.09 \\
14.5--15.0 &  779 & 0.2300--0.1740 & 0.2020 (0.1775--0.2140) & 158.64 &  13.63 &  12.59 & 2832.91 &  243.31 &  224.74 &  3.12 & 0.08 & 0.08 \\
15.0--15.5 &  848 & 0.1740--0.1350 & 0.1545 (0.1365--0.1655) & 139.02 &  12.82 &  11.78 & 3564.67 &  328.78 &  302.06 &  3.10 & 0.09 & 0.09 \\
15.5--16.0 &  863 & 0.1350--0.1070 & 0.1210 (0.1073--0.1305) &  94.90 &  10.78 &   9.73 & 3389.36 &  385.01 &  347.46 &  2.97 & 0.11 & 0.11 \\
16.0--16.5 &  931 & 0.1070--0.0862 & 0.0966 (0.0858--0.1048) &  45.73 &   7.82 &   6.74 & 2198.46 &  375.84 &  324.22 &  2.69 & 0.16 & 0.16 \\
16.5--17.0 &  953 & 0.0862--0.0710 & 0.0786 (0.0671--0.0855) &  38.41 &   7.26 &   6.18 & 2527.30 &  477.51 &  406.43 &  2.66 & 0.17 & 0.18 \\
17.0--17.5 &  747 & 0.0710--0.0549 & 0.0630 (0.0534--0.0711) &  35.81 &   7.05 &   5.96 & 2224.41 &  437.69 &  370.40 &  2.51 & 0.18 & 0.18 \\
17.5--18.0 &  508 & 0.0549--0.0486 & 0.0517 (0.0463--0.0589) &  16.22 &   5.12 &   4.00 & 2574.60 &  812.61 &  634.33 &  2.49 & 0.27 & 0.28 \\
18.0--18.5 &  268 & 0.0486--0.0440 & 0.0463 (0.0414--0.0506) &  16.70 &   5.18 &   4.06 & 3630.65 & 1125.53 &  881.74 &  2.59 & 0.27 & 0.28 \\
18.5--19.0 &  224 & 0.0440--0.0396 & 0.0418 (0.0383--0.0461) &  13.66 &   4.80 &   3.66 & 3105.68 & 1090.16 &  832.42 &  2.47 & 0.30 & 0.31 \\
19.0--19.5 &  164 & 0.0396--0.0368 & 0.0382 (0.0356--0.0420) &  11.34 &   4.48 &   3.33 & 4051.07 & 1599.11 & 1189.51 &  2.55 & 0.33 & 0.35 \\
19.5--20.0 &  193 & 0.0368--0.0339 & 0.0353 (0.0330--0.0390) &  15.26 &   5.00 &   3.87 & 5260.69 & 1724.40 & 1335.78 &  2.63 & 0.28 & 0.29 \\
20.0--20.5 &  223 & 0.0339--0.0311 & 0.0325 (0.0305--0.0370) &  38.49 &   7.26 &   6.18 & 13745.71 & 2594.29 & 2208.46 &  3.01 & 0.17 & 0.18 \\
20.5--21.0 &  145 & 0.0311--0.0294 & 0.0302 (0.0286--0.0350) &  15.99 &   5.09 &   3.97 & 9408.24 & 2995.26 & 2334.04 &  2.82 & 0.28 & 0.29 \\
 \hline
\end{tabular}
\end{table*}
%

%
%
\begin{table*}
  \caption{Same as Table \ref{tab_Pleiades:LF_MF_probabilistic} but
for method \#2\@.
}
  \label{tab_Pleiades:LF_MF_method2}
  \begin{tabular}{c c c c | c c c | c c c | c c c}
  \hline
Mag range &   Nb\_obj & Mass range  & Mid-mass & dN   & errH & errL & dN/dM & errH & errL & dN/dlogM & errH & errL \cr
  \hline
12.0--12.5 &   11 & 0.6200--0.5400 & 0.5800 (0.5550--0.5890) &  11.00 &   4.43 &   3.28 &  137.50 &   55.35 &   40.98 &  2.26 & 0.34 & 0.35 \\
12.5--13.0 &   54 & 0.5400--0.4690 & 0.5045 (0.4850--0.5125) &  54.00 &   8.40 &   7.33 &  760.56 &  118.30 &  103.26 &  2.95 & 0.14 & 0.15 \\
13.0--13.5 &   91 & 0.4690--0.3890 & 0.4290 (0.4025--0.4380) &  91.00 &  10.58 &   9.53 & 1137.50 &  132.23 &  119.08 &  3.05 & 0.11 & 0.11 \\
13.5--13.0 &  106 & 0.3890--0.3030 & 0.3460 (0.3145--0.3595) & 106.00 &  11.33 &  10.28 & 1232.56 &  131.77 &  119.58 &  2.99 & 0.10 & 0.10 \\
14.0--14.5 &  162 & 0.3030--0.2300 & 0.2665 (0.2365--0.2800) & 162.00 &  13.76 &  12.72 & 2219.18 &  188.46 &  174.22 &  3.13 & 0.08 & 0.08 \\
14.5--15.0 &  187 & 0.2300--0.1740 & 0.2020 (0.1775--0.2140) & 187.00 &  14.70 &  13.67 & 3339.29 &  262.54 &  244.03 &  3.19 & 0.08 & 0.08 \\
15.0--15.5 &  160 & 0.1740--0.1350 & 0.1545 (0.1365--0.1655) & 160.00 &  13.68 &  12.64 & 4102.56 &  350.74 &  324.08 &  3.16 & 0.08 & 0.08 \\
15.5--16.0 &   97 & 0.1350--0.1070 & 0.1210 (0.1073--0.1305) &  97.00 &  10.89 &   9.84 & 3464.29 &  388.82 &  351.29 &  2.98 & 0.11 & 0.11 \\
16.0--16.5 &   69 & 0.1070--0.0862 & 0.0966 (0.0858--0.1048) &  69.00 &   9.35 &   8.29 & 3317.31 &  449.60 &  398.63 &  2.87 & 0.13 & 0.13 \\
16.5--17.0 &   58 & 0.0862--0.0710 & 0.0786 (0.0671--0.0855) &  58.00 &   8.66 &   7.60 & 3815.79 &  570.06 &  499.96 &  2.84 & 0.14 & 0.14 \\
17.0--17.5 &   46 & 0.0710--0.0549 & 0.0630 (0.0534--0.0711) &  46.00 &   7.84 &   6.76 & 2857.14 &  486.79 &  420.12 &  2.61 & 0.16 & 0.16 \\
17.5--18.0 &   20 & 0.0549--0.0486 & 0.0517 (0.0463--0.0589) &  20.00 &   5.56 &   4.44 & 3174.60 &  881.78 &  705.41 &  2.58 & 0.25 & 0.25 \\
18.0--18.5 &   18 & 0.0486--0.0440 & 0.0463 (0.0414--0.0506) &  18.00 &   5.33 &   4.21 & 3913.04 & 1158.72 &  915.89 &  2.62 & 0.26 & 0.27 \\
18.5--19.0 &   18 & 0.0440--0.0396 & 0.0418 (0.0383--0.0461) &  18.00 &   5.33 &   4.21 & 4090.91 & 1211.39 &  957.52 &  2.59 & 0.26 & 0.27 \\
19.0--19.5 &    9 & 0.0396--0.0368 & 0.0382 (0.0356--0.0420) &   9.00 &   4.12 &   2.96 & 3214.29 & 1472.32 & 1056.44 &  2.45 & 0.38 & 0.40 \\
19.5--20.0 &   15 & 0.0368--0.0339 & 0.0353 (0.0330--0.0390) &  15.00 &   4.97 &   3.84 & 5172.41 & 1713.32 & 1324.34 &  2.62 & 0.29 & 0.30 \\
20.0--20.5 &   22 & 0.0339--0.0311 & 0.0325 (0.0305--0.0370) &  22.00 &   5.77 &   4.66 & 7857.14 & 2060.61 & 1665.60 &  2.77 & 0.23 & 0.24 \\
20.5--21.0 &    4 & 0.0311--0.0294 & 0.0302 (0.0286--0.0350) &   4.00 &   3.18 &   1.94 & 2352.94 & 1870.26 & 1139.11 &  2.21 & 0.58 & 0.66 \\
 \hline
\end{tabular}
\end{table*}

Assuming that the observed lithium depletion boundary is at 
M $\sim$ 0.075 M$_{\odot}$ \citep[M$_{Z}$ = 11.44][]{stauffer98,barrado04b} 
and a distance of 120.2 pc, the sample extracted by method \#2 contains 1147 
Pleiades member candidates, divided up into 978 stars (83.3\%) and 169 brown 
dwarfs (14.7\%). Similar percentages are obtained considering the sample of 947 
high probability members (p$\geq$60\%) identified in the probabilistic approach 
with 10.3\% of brown dwarfs. Hence, the star ($\sim$0.6--0.08 M$_{\odot}$) to 
brown dwarf (0.08--0.03 M$_{\odot}$) ratio in the Pleiades lies between 
5.4--6.3 and 8.6--9.3 if we consider a 3$\sigma$ limit in the distance of the 
Pleiades \citep[114.5--125.9 pc;][]{vanLeeuwen09}. These numbers are in 
agreement with measurements (with slightly different mass ranges depending on 
the survey) derived from the field mass function 
\citep[1.7--5.3;][]{kroupa02,chabrier05a,andersen06}, young star-forming regions 
\citep[3.0--6.4 for the Trapezium Cluster; 3.8-4.3 for $\sigma$ Orionis; 3.8 
for Chamaeleon; 8.3--11.6 for IC\,348;][]{hillenbrand00,muench02,luhman03b,andersen06,luhman07d,lodieu09e}, 
open clusters \citep[3.7 for the Pleiades and 4.5 for M35;][]{bouvier98,barrado01a},
and hydrodynamical simulations of star clusters \citep[3.8--5.0;][]{bate09,bate11b}.

\subsection{The cluster mass function}
\label{Pleiades:IMF_MF}

In this section we adopt the logarithmic form of the Initial Mass Function as
originally proposed by \citet{salpeter55}:
$\xi$($\log_{10}m$) = d$n$/d$\log_{10}$($m$) $\propto$ m$^{-\alpha}$.
We converted the luminosity into a mass function using the NextGen models 
\citep{baraffe98} for stars and brown dwarfs more massive than 50 M$_{\rm Jup}$ 
(T$_{\rm eff}$) and the DUSTY models \citep{chabrier00c} for less massive brown 
dwarfs. The $Z$ = 12--21.5 mag range translates into masses between 0.62 and 
0.03 M$_{\odot}$, assuming a revised distance of 120.2 pc \citep{vanLeeuwen09} 
and an age of 120 Myr for which the models are computed.

\begin{figure}
   \includegraphics[width=0.75\linewidth,angle=270]{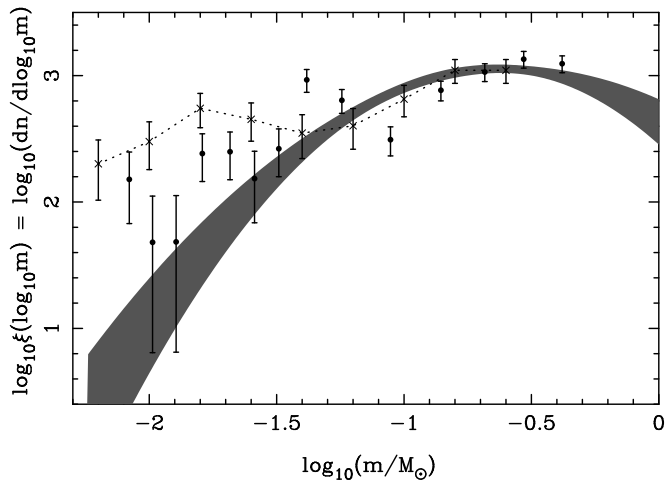}
   \caption{Mass function comparison between the Pleiades (shaded region showing 
a $1\sigma$ confidence interval from a weighted least--squares log--normal fit 
to the data presented here and extrapolated to lower masses) along with data from 
$\sigma$ Orionis \citep[filled circles;][]{lodieu09e} and Upper Sco
\citep[crosses joined by a dashed line;][]{lodieu11a}. These have been normalised 
at the peak of the MF\@.} 
   \label{fig_Pleiades:imf_compare}
\end{figure}

We included in Fig.\ \ref{fig_Pleiades:LF_and_MF} errors in both the
x-axis ($\log$M) and y-axis (dN/d$\log$M) as follows. For the error bars
on the masses, we considered three times the uncertainties on the age 
\citep[125$\pm$8 Myr;][]{stauffer98} and distance 
\citep[120.2$\pm$1.9 pc;][]{vanLeeuwen09} of the Pleiades given us a validity
range of 3$\sigma$ on the x-axis.
Hence, we computed the masses with the 100 Myr NextGen and DUSTY isochrones
shifted at a distance of 114.5 pc to define the lower limit and repeated the
procedure with the 150 Myr isochrones for a distance of 125.9 pc as upper limits.
(The uncertainties on the y-axis i.e.\ the dN/d$\log$M values are simply
Gehrels error bars). The highest-mass point is incomplete due to the saturation
of the GCS as are the two lowest-mass points of the mass function because
the last two magnitude bins in the $Z$-band are more contaminated. Using these 
upper and lower bounds for the predicted
masses for cluster members we refit the log--normal MF to examine the
effects on the parameters of the fit (Fig.\ \ref{fig_Pleiades:imf_compare}.) 
Following the \citet{chabrier03} Gaussian parameterisation (his Equation~12) 
4e find a characteristic mass $m_c=0.24^{+0.01}_{-0.03}$~M$_{\odot}$ and a 
mass dispersion $\sigma=0.44\pm0.01$ (\citet{chabrier05a} quotes $m_c=0.25$ 
and $\sigma=0.55$ for the disk system MF whereas \citet{chabrier03} quotes
$m_c=0.22$ and $\sigma=0.57$).

Overall we find that our Pleiades mass function is well represented by a
log-normal form over the 0.6--0.03 M$_{\odot}$ mass range with a characteristic
mass of 0.24 M$_{\odot}$ (Fig.\ \ref{fig_Pleiades:imf_compare}). 
This result is in agreement with all 
previous studies in the Pleiades \citep{moraux01,tej02,dobbie02a,deacon04} 
over the same mass range and consistent with the extrapolation of the system 
field mass function \citep{chabrier05a} which can also reproduce 
preliminary densities of field L and T dwarfs found in large-scale surveys 
\citep{metchev08,burningham10b,reyle10,kirkpatrick11} as displayed in 
Fig.\ \ref{fig_Pleiades:LF_and_MF}. All these determinations of the mass 
function support the universality of the IMF as discussed in the review of 
\citet{bastian10} except for the case of Taurus \citep{briceno02,luhman03a}. 
The latest hydrodynamical simulation of \citet{bate11b} is able to reproduce 
the observed field mass function \citep{kroupa02,chabrier05a} with high 
confidence after inclusion of radiative feedback, in agreement with 
independent calculations \citep{offner09,urban10,krumholz11}.

%
%

\begin{figure}
   \includegraphics[width=0.70\linewidth,angle=270]{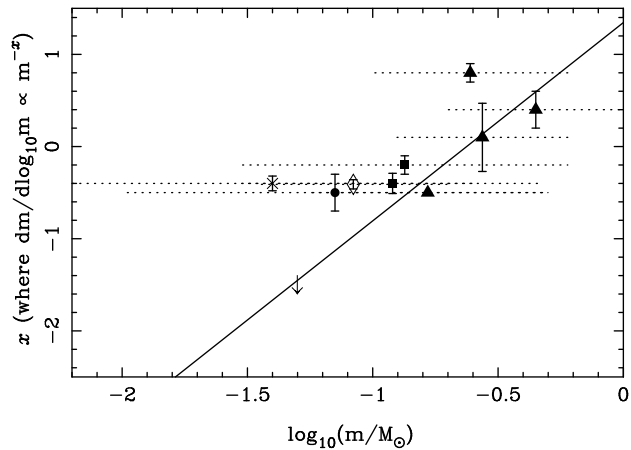}
   \caption{'$x$-plot' equivalent to the `Alpha plot' (where $x=\alpha-1$, 
cf.\ \citet{chabrier05a}) showing 
the gradient of the mass function as a function of mass. The straight line is 
the gradient of the log--normal fit to the Pleiades data presented herein, 
while the filled circle is the recent GCS results from $\sigma$ Orionis 
\citep[$\alpha$=0.5$\pm$0.2;][]{lodieu09e}, the cross is from a power-law fit 
to the most recent Upper Sco results \citep[$x=-0.6\pm0.08$;][]{lodieu11a},
the open diamond for Alpha Per ($\alpha$=0.59$\pm$0.05) from \citet{barrado02a},
and the filled triangles are the results from Praesepe 
\citep[$\alpha$=1.4$\pm$0.2; $\alpha$=1.8$\pm$0.1; $\alpha$=1.11$\pm$0.37;][]{kraus07d,boudreault10,baker10}. 
The arrow indicates the upper limit to the mass function of the field brown 
dwarf population \citep{burningham10b}. Horizontal dotted lines indicate 
mass ranges for power--law IMF determinations.} 
   \label{fig:alpha_plot}
\end{figure}

It is interesting to compare an extrapolation to lower masses of the 
log--normal Pleiades mass function derived here with results from other 
younger GCS targets (Fig.\ \ref{fig_Pleiades:imf_compare}).
While the Pleiades is measured in a higher mass range and shows a closely
log--normal form, the IMF from a carefully cleaned spectroscopic sample in 
Upper Sco \citep{lodieu11a} penetrates to lower masses and seems to be much 
shallower in the substellar regime, implying the presence of more brown dwarfs. 
Results for the sigma Orionis cluster show a similar trend~\citep{lodieu09e}, 
although to a less obvious extent. Of course, these mass functions have been 
derived assuming that the evolutionary models accurately predict colours, 
bolometric magnitudes and temperatures at the different ages (especially 
ages $<$\,10 Myr) and assuming also that any systematic errors introduced by 
not accounting for unresolved binarity cancel in the comparison. 
In Fig.\ \ref{fig:alpha_plot} we show an '$x$-plot' equivalent to the
`alpha plot' with $x$ = $\alpha$ $-$ 1), i.e.\ a plot of the gradient of 
mass function as a function of mass, in order to compare with the constraints 
set on the field IMF gradient by the UKIDSS Large Area Survey T--type brown 
dwarf searches \citep[e.g.][]{lodieu07b,pinfield08,burningham10b}). 
While the mass functions of the more aged populations like Praesepe 
\citep{kraus07d,boudreault10,baker10} or the field
\citep{metchev08,burningham10b,reyle10,kirkpatrick11}
show gradients consistent with the Pleiades log--normal, power--law fits to 
the very young Upper Sco \citep[$x=-0.4\pm0.08$;][]{lodieu11a} and $\sigma$ 
Orionis \citep[$x$=-0.5$\pm$0.2;][]{lodieu09e} clusters are flatter. 
This may be evidence of a variation in the substellar IMF, or simply an 
artefact of systematic errors in the evolutionary model predictions at very 
young ages or age spreads. We should emphasise that the upper limit
set for the field mass function by \citet{burningham10b} remains under
debate (due to the lack of constraints on masses and ages) as the CFHT brown 
dwarf survey \citep{reyle10} and the preliminary densities determined by 
WISE \citep{kirkpatrick11} suggest positive $\alpha$ values for the mass 
function in the T dwarf regime \citep[see also][]{metchev08}.

%
%

%
%
\section{Summary}
\label{Pleiades:summary}

We have presented the outcome of a wide ($\sim$80 square degrees) and deep 
($J$ $\sim$ 18.8 mag) survey in the Pleiades open cluster as part of the UKIDSS 
Galactic Clusters Survey Data Release 9\@. The main results of our analysis
can be summarised as follows:
\begin{itemize}
\item we recovered Pleiades member candidates previously published
and updated their membership assignations
\item we selected photometrically and astrometrically potential Pleiades
member candidates using two independent but complementary methods: the
probabilistic analysis and a more standard method combining photometry and proper
motion cuts
\item we derived a BD binary fraction around 25.6$\pm$4.5\% in the 
0.075--0.03 M$_{\odot}$ mass range with a difference of a factor of two between 
high-mass (0.075--0.05 M$_{\odot}$) and low-mass (0.05--0.03 M$_{\odot}$)
brown dwarfs for separations less than $\sim$100--200 au
\item we investigated the $K$-band variability of Pleiades members and
found virtually no variability at the level of 0.08 mag
\item we derived the luminosity function from both selection methods and
found no difference within the error bars
\item we derived the Pleiades mass function which is best fit by a 
log-normal function peaking at 0.16--0.20 M$_{\odot}$ in the 0.6--0.03 M$_{\odot}$ 
range, in agreement with previous studies in the cluster and the extrapolation
of the system field mass function
\end{itemize}

This paper represents a significant improvement in our knowledge of the Pleiades 
population and the cluster mass function in the substellar regime over the full 
cluster. We believe that this paper will represent a reference for many more years
to come. We will now extend this study to other regions
surveyed by the GCS to adress the question of the universality of the
mass function using an homogeneous set of photometric and astrometric data.
Future work to constraint current models of star formation include the search
for companions to investigate their multiplicity properties, the
determination of the radial velocities of Pleiades members, and deeper surveys
to test the theory of the fragmentation limit.

%
%
\section*{Acknowledgments}

NL is funded by the Ram\'on y Cajal fellowship number 08-303-01-02 and the 
national program AYA2010-19136 funded by the Spanish ministry of science and 
innovation. This work is based in part on data obtained as part of the UKIRT 
Infrared Deep Sky Survey (UKIDSS). The UKIDSS project is defined in 
\citet{lawrence07}. UKIDSS uses the UKIRT Wide Field Camera
\citep[WFCAM;][]{casali07}. The photometric system is described in
\citet{hewett06}, and the calibration is described in \citet{hodgkin09}.
The pipeline processing and science archive are described in Irwin et al.\
(in prep) and \citet{hambly08}, respectively.
We thank our colleagues at the UK Astronomy Technology Centre, the Joint 
Astronomy Centre in Hawaii, the Cambridge Astronomical Survey and Edinburgh 
Wide Field Astronomy Units for building and operating WFCAM and its 
associated data flow system.
We are grateful to Isabelle Baraffe and France Allard for 
providing us with the NextGen, DUSTY and COND models for the WFCAM filters.

This research has made use of the Simbad database, operated at the Centre de 
Donn\'ees Astronomiques de Strasbourg (CDS), and of NASA's Astrophysics Data 
System Bibliographic Services (ADS). This publication has also made use of data 
products from the Two Micron All Sky Survey, which is a joint project of the
University of Massachusetts and the Infrared Processing and Analysis 
Center/California Institute of Technology, funded by the National Aeronautics 
and Space Administration and the National Science Foundation.

%
%
\bibliographystyle{mn2e}
\bibliography{../../AA/mnemonic,../../AA/biblio_old}

\begin{thebibliography}{}

\bibitem[\protect\citeauthoryear{{Adams}, {Stauffer}, {Monet}, {Skrutskie} \&
  {Beichman}}{{Adams} et~al.}{2001}]{adams01a}
{Adams} J.~D.,  {Stauffer} J.~R.,  {Monet} D.~G.,  {Skrutskie} M.~F.,
  {Beichman} C.~A.,  2001, AJ, 121, 2053

\bibitem[\protect\citeauthoryear{{Andersen}, {Meyer}, {Oppenheimer}, {Dougados}
  \& {Carpenter}}{{Andersen} et~al.}{2006}]{andersen06}
{Andersen} M.,  {Meyer} M.~R.,  {Oppenheimer} B.,  {Dougados} C.,
  {Carpenter} J.,  2006, AJ, 132, 2296

\bibitem[\protect\citeauthoryear{{Artyukhina}}{{Artyukhina}}{1969}]{artyukhina69}
{Artyukhina} N.~M.,  1969, 12, 987

\bibitem[\protect\citeauthoryear{{Baker}, {Jameson}, {Casewell}, {Deacon},
  {Lodieu} \& {Hambly}}{{Baker} et~al.}{2010}]{baker10}
{Baker} D.~E.~A.,  {Jameson} R.~F.,  {Casewell} S.~L.,  {Deacon} N.,  {Lodieu}
  N.,    {Hambly} N.,  2010, MNRAS, 408, 2457

\bibitem[\protect\citeauthoryear{{Baraffe}, {Chabrier}, {Allard} \&
  {Hauschildt}}{{Baraffe} et~al.}{1998}]{baraffe98}
{Baraffe} I.,  {Chabrier} G.,  {Allard} F.,    {Hauschildt} P.~H.,  1998, A\&A,
  337, 403

\bibitem[\protect\citeauthoryear{{Barrado y Navascu{\' e}s}, {Stauffer} \&
  {Jayawardhana}}{{Barrado y Navascu{\' e}s} et~al.}{2004}]{barrado04b}
{Barrado y Navascu{\' e}s} D.,  {Stauffer} J.~R.,    {Jayawardhana} R.,  2004,
  ApJ, 614, 386

\bibitem[\protect\citeauthoryear{{Barrado y Navascu{\'e}s}, {Bouvier},
  {Stauffer}, {Lodieu} \& {McCaughrean}}{{Barrado y Navascu{\'e}s}
  et~al.}{2002}]{barrado02a}
{Barrado y Navascu{\'e}s} D.,  {Bouvier} J.,  {Stauffer} J.~R.,  {Lodieu} N.,
   {McCaughrean} M.~J.,  2002, A\&A, 395, 813

\bibitem[\protect\citeauthoryear{{Barrado y Navascu{\'e}s}, {Stauffer},
  {Bouvier} \& {Mart{\'{\i}}n}}{{Barrado y Navascu{\'e}s}
  et~al.}{2001}]{barrado01a}
{Barrado y Navascu{\'e}s} D.,  {Stauffer} J.~R.,  {Bouvier} J.,
  {Mart{\'{\i}}n} E.~L.,  2001, ApJ, 546, 1006

\bibitem[\protect\citeauthoryear{{Basri} \& {Mart{\'{\i}}n}}{{Basri} \&
  {Mart{\'{\i}}n}}{1999}]{basri99b}
{Basri} G.,  {Mart{\'{\i}}n} E.~L.,  1999, AJ, 118, 2460

\bibitem[\protect\citeauthoryear{{Basri} \& {Reiners}}{{Basri} \&
  {Reiners}}{2006}]{basri06}
{Basri} G.,  {Reiners} A.,  2006, AJ, 132, 663

\bibitem[\protect\citeauthoryear{{Bastian}, {Covey} \& {Meyer}}{{Bastian}
  et~al.}{2010}]{bastian10}
{Bastian} N.,  {Covey} K.~R.,    {Meyer} M.~R.,  2010, ArXiv e-prints

\bibitem[\protect\citeauthoryear{{Bate}}{{Bate}}{2009}]{bate09}
{Bate} M.~R.,  2009, MNRAS, 392, 590

\bibitem[\protect\citeauthoryear{{Bate}}{{Bate}}{2011}]{bate11b}
{Bate} M.~R.,  2011, MNRAS, 418, 703

\bibitem[\protect\citeauthoryear{{B{\'e}jar}, {Mart{\'{\i}}n}, {Zapatero
  Osorio}, {Rebolo}, {Barrado y Navascu{\' e}s}, {Bailer-Jones}, {Mundt},
  {Baraffe}, {Chabrier} \& {Allard}}{{B{\'e}jar} et~al.}{2001}]{bejar01}
{B{\'e}jar} V.~J.~S.,  et al.\ 2001, ApJ, 556, 830

\bibitem[\protect\citeauthoryear{{Bihain}, {Rebolo}, {B{\'e}jar}, {Caballero},
  {Bailer-Jones}, {Mundt}, {Acosta-Pulido} \& {Manchado Torres}}{{Bihain}
  et~al.}{2006}]{bihain06}
{Bihain} G.,  {Rebolo} R.,  {B{\'e}jar} V.~J.~S.,  {Caballero} J.~A.,
  {Bailer-Jones} C.~A.~L.,  {Mundt} R.,  {Acosta-Pulido} J.~A.,    {Manchado
  Torres} A.,  2006, A\&A, 458, 805

\bibitem[\protect\citeauthoryear{{Bihain}, {Rebolo}, {Zapatero Osorio},
  {B{\'e}jar} \& {Caballero}}{{Bihain} et~al.}{2010}]{bihain10a}
{Bihain} G.,  {Rebolo} R.,  {Zapatero Osorio} M.~R.,  {B{\'e}jar} V.~J.~S.,
  {Caballero} J.~A.,  2010, A\&A, 519, A93

\bibitem[\protect\citeauthoryear{{Boudreault}, {Bailer-Jones}, {Goldman},
  {Henning} \& {Caballero}}{{Boudreault} et~al.}{2010}]{boudreault10}
{Boudreault} S.,  {Bailer-Jones} C.~A.~L.,  {Goldman} B.,  {Henning} T.,
  {Caballero} J.~A.,  2010, A\&A, 510, A27+

\bibitem[\protect\citeauthoryear{{Bouvier}, {Stauffer}, {Mart\'{\i}n}, {Barrado
  y Navascu\'es}, {Wallace} \& {B\'ejar}}{{Bouvier} et~al.}{1998}]{bouvier98}
{Bouvier} J.,  {Stauffer} J.~R.,  {Mart\'{\i}n} E.~L.,  {Barrado y Navascu\'es}
  D.,  {Wallace} B.,    {B\'ejar} V.~J.~S.,  1998, A\&A, 336, 490

\bibitem[\protect\citeauthoryear{{Bouy}, {Moraux}, {Bouvier}, {Brandner},
  {Mart{\'{\i}}n}, {Allard}, {Baraffe} \& {Fern{\'a}ndez}}{{Bouy}
  et~al.}{2006}]{bouy06a}
{Bouy} H.,  {Moraux} E.,  {Bouvier} J.,  {Brandner} W.,  {Mart{\'{\i}}n} E.~L.,
   {Allard} F.,  {Baraffe} I.,    {Fern{\'a}ndez} M.,  2006, ApJ, 637, 1056

\bibitem[\protect\citeauthoryear{{Brice{\~ n}o}, {Luhman}, {Hartmann},
  {Stauffer} \& {Kirkpatrick}}{{Brice{\~ n}o} et~al.}{2002}]{briceno02}
{Brice{\~ n}o} C.,  {Luhman} K.~L.,  {Hartmann} L.,  {Stauffer} J.~R.,
  {Kirkpatrick} J.~D.,  2002, ApJ, 580, 317

\bibitem[\protect\citeauthoryear{{Burgasser}, {Reid}, {Siegler}, {Close},
  {Allen}, {Lowrance} \& {Gizis}}{{Burgasser} et~al.}{2007}]{burgasser07a}
{Burgasser} A.~J.,  et al.\ eds, Protostars and Planets V {Not Alone: Tracing 
the Origins of Very-Low-Mass Stars and Brown Dwarfs Through Multiplicity Studies}.
pp 427--441

\bibitem[\protect\citeauthoryear{{Burningham}, {Pinfield}, {Lucas}, {Leggett},
  {Deacon}, {Tamura}, {Tinney}, {Lodieu} \& {11 co-authors}}{{Burningham}
  et~al.}{2010}]{burningham10b}
{Burningham} B., et al.\ 2010, MNRAS, 406, 1885

\bibitem[\protect\citeauthoryear{{Casali}, {Adamson}, {Alves de Oliveira},
  {Almaini}, {Burch}, {Chuter}, {Elliot} \& {23 co-authors}}{{Casali}
  et~al.}{2007}]{casali07}
{Casali} M.,  et al.\ 2007, A\&A, 467, 777

\bibitem[\protect\citeauthoryear{{Casewell}, {Dobbie}, {Hodgkin}, {Moraux},
  {Jameson}, {Hambly}, {Irwin} \& {Lodieu}}{{Casewell}
  et~al.}{2007}]{casewell07}
{Casewell} S.~L.,  {Dobbie} P.~D.,  {Hodgkin} S.~T.,  {Moraux} E.,  {Jameson}
  R.~F.,  {Hambly} N.~C.,  {Irwin} J.,    {Lodieu} N.,  2007, MNRAS, 378, 1131

\bibitem[\protect\citeauthoryear{{Casewell}, {Dobbie}, {Hodgkin}, {Moraux},
  {Jameson}, {Hambly}, {Irwin} \& {Lodieu}}{{Casewell}
  et~al.}{2010}]{casewell10}
{Casewell} S.~L.,  {Dobbie} P.~D.,  {Hodgkin} S.~T.,  {Moraux} E.,  {Jameson}
  R.~F.,  {Hambly} N.~C.,  {Irwin} J.,    {Lodieu} N.,  2010, MNRAS, 402, 1407

\bibitem[\protect\citeauthoryear{{Casewell}, {Jameson}, {Burleigh}, {Dobbie},
  {Roy}, {Hodgkin} \& {Moraux}}{{Casewell} et~al.}{2011}]{casewell11}
{Casewell} S.~L.,  {Jameson} R.~F.,  {Burleigh} M.~R.,  {Dobbie} P.~D.,  {Roy}
  M.,  {Hodgkin} S.~T.,    {Moraux} E.,  2011, MNRAS, 412, 2071

\bibitem[\protect\citeauthoryear{{Chabrier}}{{Chabrier}}{2003}]{chabrier03}
{Chabrier} G.,  2003, PASP, 115, 763

\bibitem[\protect\citeauthoryear{{Chabrier}}{{Chabrier}}{2005}]{chabrier05a}
{Chabrier} G.,  2005, in {E.~Corbelli, F.~Palla, \& H.~Zinnecker} ed., The
  Initial Mass Function 50 Years Later Vol.~327 of Astrophysics and Space
  Science Library, {The Initial Mass Function: from Salpeter 1955 to 2005}.
p.~41

\bibitem[\protect\citeauthoryear{{Chabrier}, {Baraffe}, {Allard} \&
  {Hauschildt}}{{Chabrier} et~al.}{2000}]{chabrier00c}
{Chabrier} G.,  {Baraffe} I.,  {Allard} F.,    {Hauschildt} P.,  2000, ApJ,
  542, 464

\bibitem[\protect\citeauthoryear{{Cutri}, {Skrutskie}, {van Dyk}, {Beichman},
  {Carpenter}, {Chester}, {Cambresy}, {Evans}, {Fowler}, {Gizis} \& {15
  coauthors}}{{Cutri} et~al.}{2003}]{cutri03}
{Cutri} R.~M., et al.\ 2003, 2MASS All Sky Catalog of point sources, 2246

\bibitem[\protect\citeauthoryear{{Deacon} \& {Hambly}}{{Deacon} \&
  {Hambly}}{2004}]{deacon04}
{Deacon} N.~R.,  {Hambly} N.~C.,  2004, A\&A, 416, 125

\bibitem[\protect\citeauthoryear{{Dobbie}, {Kenyon}, {Jameson}, {Hodgkin},
  {Pinfield} \& {Osborne}}{{Dobbie} et~al.}{2002}]{dobbie02a}
{Dobbie} P.~D.,  {Kenyon} F.,  {Jameson} R.~F.,  {Hodgkin} S.~T.,  {Pinfield}
  D.~J.,    {Osborne} S.~L.,  2002, MNRAS, 335, 687

\bibitem[\protect\citeauthoryear{{Dobbie}, {Pinfield}, {Jameson} \&
  {Hodgkin}}{{Dobbie} et~al.}{2002}]{dobbie02b}
{Dobbie} P.~D.,  {Pinfield} D.~J.,  {Jameson} R.~F.,    {Hodgkin} S.~T.,  2002,
  MNRAS, 335, L79

\bibitem[\protect\citeauthoryear{{Festin}}{{Festin}}{1998}]{festin98}
{Festin} L.,  1998, A\&A, 333, 497

\bibitem[\protect\citeauthoryear{{Gatewood}, {de Jonge} \& {Han}}{{Gatewood}
  et~al.}{2000}]{gatewood00}
{Gatewood} G.,  {de Jonge} J.~K.,    {Han} I.,  2000, ApJ, 533, 938

\bibitem[\protect\citeauthoryear{{Hambly}, {Collins}, {Cross}, {Mann}, {Read},
  {Sutorius}, {Bond}, {Bryant}, {Emerson}, {Lawrence}, {Rimoldini}, {Stewart},
  {Williams}, {Adamson}, {Hirst}, {Dye} \& {Warren}}{{Hambly}
  et~al.}{2008}]{hambly08}
{Hambly} N.~C., et al.\ 2008, MNRAS, 384, 637

\bibitem[\protect\citeauthoryear{{Hambly}, {Hawkins} \& {Jameson}}{{Hambly}
  et~al.}{1993}]{hambly93}
{Hambly} N.~C.,  {Hawkins} M.~R.~S.,    {Jameson} R.~F.,  1993, A\&AS, 100, 607

\bibitem[\protect\citeauthoryear{{Hambly}, {Hodgkin}, {Cossburn} \&
  {Jameson}}{{Hambly} et~al.}{1999}]{hambly99}
{Hambly} N.~C.,  {Hodgkin} S.~T.,  {Cossburn} M.~R.,    {Jameson} R.~F.,  1999,
  MNRAS, 303, 835

\bibitem[\protect\citeauthoryear{{Hambly}, {Steele}, {Hawkins} \&
  {Jameson}}{{Hambly} et~al.}{1995}]{hambly95}
{Hambly} N.~C.,  {Steele} I.~A.,  {Hawkins} M.~R.~S.,    {Jameson} R.~F.,
  1995, MNRAS, 273, 505

\bibitem[\protect\citeauthoryear{{Haro}, {Chavira} \& {Gonzalez}}{{Haro}
  et~al.}{1982}]{haro82}
{Haro} G.,  {Chavira} E.,    {Gonzalez} G.,  1982, Boletin del Instituto de
  Tonantzintla, 3, 3

\bibitem[\protect\citeauthoryear{{Hertzsprung}}{{Hertzsprung}}{1947}]{hertzsprung47}
{Hertzsprung} E.,  1947, Annalen van de Sterrewacht te Leiden, 19, A1

\bibitem[\protect\citeauthoryear{{Hewett}, {Warren}, {Leggett} \&
  {Hodgkin}}{{Hewett} et~al.}{2006}]{hewett06}
{Hewett} P.~C.,  {Warren} S.~J.,  {Leggett} S.~K.,    {Hodgkin} S.~T.,  2006,
  MNRAS, 367, 454

\bibitem[\protect\citeauthoryear{{Hillenbrand} \& {Carpenter}}{{Hillenbrand} \&
  {Carpenter}}{2000}]{hillenbrand00}
{Hillenbrand} L.~A.,  {Carpenter} J.~M.,  2000, ApJ, 540, 236

\bibitem[\protect\citeauthoryear{{Hodgkin}, {Irwin}, {Hewett} \&
  {Warren}}{{Hodgkin} et~al.}{2009}]{hodgkin09}
{Hodgkin} S.~T.,  {Irwin} M.~J.,  {Hewett} P.~C.,    {Warren} S.~J.,  2009,
  MNRAS, 394, 675

\bibitem[\protect\citeauthoryear{{Jameson}, {Dobbie}, {Hodgkin} \&
  {Pinfield}}{{Jameson} et~al.}{2002}]{jameson02}
{Jameson} R.~F.,  {Dobbie} P.~D.,  {Hodgkin} S.~T.,    {Pinfield} D.~J.,  2002,
  MNRAS, 335, 853

\bibitem[\protect\citeauthoryear{{Jameson} \& {Skillen}}{{Jameson} \&
  {Skillen}}{1989}]{jameson89}
{Jameson} R.~F.,  {Skillen} I.,  1989, MNRAS, 239, 247

\bibitem[\protect\citeauthoryear{{Johnson}}{{Johnson}}{1957}]{johnson57}
{Johnson} H.~L.,  1957, ApJ, 126, 121

\bibitem[\protect\citeauthoryear{{Jones}}{{Jones}}{1981}]{jones81}
{Jones} B.~F.,  1981, AJ, 86, 290

\bibitem[\protect\citeauthoryear{{Jones} \& {Stauffer}}{{Jones} \&
  {Stauffer}}{1991}]{jones91}
{Jones} B.~F.,  {Stauffer} J.~R.,  1991, AJ, 102, 1080

\bibitem[\protect\citeauthoryear{{Kirkpatrick}, {Cushing}, {Gelino},
  {Griffith}, {Skrutskie}, {Marsh}, {Wright}, {Mainzer}, {Eisenhardt}, {McLean}
  \& {30 co-authors}}{{Kirkpatrick} et~al.}{2011}]{kirkpatrick11}
{Kirkpatrick} J.~D., et al.\ 2011, ApJS, 197, 17

\bibitem[\protect\citeauthoryear{{Kraus} \& {Hillenbrand}}{{Kraus} \&
  {Hillenbrand}}{2007}]{kraus07d}
{Kraus} A.~L.,  {Hillenbrand} L.~A.,  2007, AJ, 134, 2340

\bibitem[\protect\citeauthoryear{{Kroupa}}{{Kroupa}}{2002}]{kroupa02}
{Kroupa} P.,  2002, Science, 295, 82

\bibitem[\protect\citeauthoryear{{Krumholz}, {Klein} \& {McKee}}{{Krumholz}
  et~al.}{2011}]{krumholz11}
{Krumholz} M.~R.,  {Klein} R.~I.,    {McKee} C.~F.,  2011, ApJ, 740, 74

\bibitem[\protect\citeauthoryear{{Lawrence}, {Warren}, {Almaini}, {Edge},
  {Hambly} \& {17 co-authors}}{{Lawrence} et~al.}{2007}]{lawrence07}
{Lawrence} A., et al.\ 2007, MNRAS, 379, 1599

\bibitem[\protect\citeauthoryear{{Lodieu}, {Dobbie}, {Deacon}, {Hodgkin},
  {Hambly} \& {Jameson}}{{Lodieu} et~al.}{2007}]{lodieu07c}
{Lodieu} N.,  {Dobbie} P.~D.,  {Deacon} N.~R.,  {Hodgkin} S.~T.,  {Hambly}
  N.~C.,    {Jameson} R.~F.,  2007, MNRAS, 380, 712

\bibitem[\protect\citeauthoryear{{Lodieu}, {Pinfield}, {Leggett}, {Jameson},
  {Mortlock}, {Warren} \& {co-authors}}{{Lodieu} et~al.}{2007}]{lodieu07b}
{Lodieu} N., et al.\ 2007, MNRAS, 379, 1423

\bibitem[\protect\citeauthoryear{{Lodieu}, {Dobbie} \& {Hambly}}{{Lodieu}
  et~al.}{2011}]{lodieu11a}
{Lodieu} N.,  {Dobbie} P.~D.,    {Hambly} N.~C.,  2011, A\&A, 527, A24

\bibitem[\protect\citeauthoryear{{Lodieu}, {Hambly} \& {Jameson}}{{Lodieu}
  et~al.}{2006}]{lodieu06}
{Lodieu} N.,  {Hambly} N.~C.,    {Jameson} R.~F.,  2006, MNRAS, 373, 95

\bibitem[\protect\citeauthoryear{{Lodieu}, {Hambly}, {Jameson}, {Hodgkin},
  {Carraro} \& {Kendall}}{{Lodieu} et~al.}{2007}]{lodieu07a}
{Lodieu} N.,  {Hambly} N.~C.,  {Jameson} R.~F.,  {Hodgkin} S.~T.,  {Carraro}
  G.,    {Kendall} T.~R.,  2007, MNRAS, 374, 372

\bibitem[\protect\citeauthoryear{{Lodieu}, {Zapatero Osorio}, {Rebolo},
  {Mart{\'{\i}}n} \& {Hambly}}{{Lodieu} et~al.}{2009}]{lodieu09e}
{Lodieu} N.,  {Zapatero Osorio} M.~R.,  {Rebolo} R.,  {Mart{\'{\i}}n} E.~L.,
  {Hambly} N.~C.,  2009, A\&A, 505, 1115

\bibitem[\protect\citeauthoryear{{Lucas} \& {Roche}}{{Lucas} \&
  {Roche}}{2000}]{lucas00}
{Lucas} P.~W.,  {Roche} P.~F.,  2000, MNRAS, 314, 858

\bibitem[\protect\citeauthoryear{{Luhman}}{{Luhman}}{1999}]{luhman99a}
{Luhman} K.~L.,  1999, ApJ, 525, 466

\bibitem[\protect\citeauthoryear{{Luhman}}{{Luhman}}{2007}]{luhman07d}
{Luhman} K.~L.,  2007, ApJS, 173, 104

\bibitem[\protect\citeauthoryear{{Luhman}, {Brice{\~ n}o}, {Stauffer},
  {Hartmann}, {Barrado y Navascu{\' e}s} \& {Caldwell}}{{Luhman}
  et~al.}{2003}]{luhman03a}
{Luhman} K.~L.,  {Brice{\~ n}o} C.,  {Stauffer} J.~R.,  {Hartmann} L.,
  {Barrado y Navascu{\' e}s} D.,    {Caldwell} N.,  2003, ApJ, 590, 348

\bibitem[\protect\citeauthoryear{{Luhman}, {Liebert} \& {Rieke}}{{Luhman}
  et~al.}{1997}]{luhman97}
{Luhman} K.~L.,  {Liebert} J.,    {Rieke} G.~H.,  1997, ApJL, 489, L165

\bibitem[\protect\citeauthoryear{{Luhman}, {Stauffer}, {Muench}, {Rieke},
  {Lada}, {Bouvier} \& {Lada}}{{Luhman} et~al.}{2003}]{luhman03b}
{Luhman} K.~L.,  {Stauffer} J.~R.,  {Muench} A.~A.,  {Rieke} G.~H.,  {Lada}
  E.~A.,  {Bouvier} J.,    {Lada} C.~J.,  2003, ApJ, 593, 1093

\bibitem[\protect\citeauthoryear{{Mart{\'{\i}}n}, {Barrado y Navascu{\' e}s},
  {Baraffe}, {Bouy} \& {Dahm}}{{Mart{\'{\i}}n} et~al.}{2003}]{martin03}
{Mart{\'{\i}}n} E.~L.,  {Barrado y Navascu{\' e}s} D.,  {Baraffe} I.,  {Bouy}
  H.,    {Dahm} S.,  2003, ApJ, 594, 525

\bibitem[\protect\citeauthoryear{{Mart\'{\i}n}, {Basri}, {Gallegos}, {Rebolo},
  {Zapatero-Osorio} \& {Bejar}}{{Mart\'{\i}n} et~al.}{1998}]{martin98b}
{Mart\'{\i}n} E.~L.,  {Basri} G.,  {Gallegos} J.~E.,  {Rebolo} R.,
  {Zapatero-Osorio} M.~R.,    {Bejar} V.~J.~S.,  1998, ApJL, 499, L61

\bibitem[\protect\citeauthoryear{{Mart{\'{\i}}n}, {Brandner}, {Bouvier},
  {Luhman}, {Stauffer}, {Basri}, {Zapatero Osorio} \& {Barrado y Navascu{\'
  e}s}}{{Mart{\'{\i}}n} et~al.}{2000}]{martin00a}
{Mart{\'{\i}}n} E.~L.,  {Brandner} W.,  {Bouvier} J.,  {Luhman} K.~L.,
  {Stauffer} J.,  {Basri} G.,  {Zapatero Osorio} M.~R.,    {Barrado y
  Navascu{\' e}s} D.,  2000, ApJ, 543, 299

\bibitem[\protect\citeauthoryear{{Mart\'{\i}n}, {Zapatero Osorio} \&
  {Rebolo}}{{Mart\'{\i}n} et~al.}{1998}]{martin98a}
{Mart\'{\i}n} E.~L.,  {Zapatero Osorio} M.~R.,    {Rebolo} R.,  1998, in ASP
  Conf.\@ Ser.\@ 134: ``Brown Dwarfs and Extrasolar Planets'', eds.\@ R.
  Rebolo, E.~L. Mart\'{\i}n, and M.~R. Zapatero Osorio {The Substellar Initial
  Mass Function in the Pleiades}.
p. p 507

\bibitem[\protect\citeauthoryear{{Maxted} \& {Jeffries}}{{Maxted} \&
  {Jeffries}}{2005}]{maxted05}
{Maxted} P.~F.~L.,  {Jeffries} R.~D.,  2005, MNRAS, 362, L45

\bibitem[\protect\citeauthoryear{{Mermilliod}}{{Mermilliod}}{1981}]{mermilliod81}
{Mermilliod} J.~C.,  1981, A\&A, 97, 235

\bibitem[\protect\citeauthoryear{{Metchev}, {Kirkpatrick}, {Berriman} \&
  {Looper}}{{Metchev} et~al.}{2008}]{metchev08}
{Metchev} S.~A.,  {Kirkpatrick} J.~D.,  {Berriman} G.~B.,    {Looper} D.,
  2008, ApJ, 676, 1281

\bibitem[\protect\citeauthoryear{{Miller} \& {Scalo}}{{Miller} \&
  {Scalo}}{1979}]{miller79}
{Miller} G.~E.,  {Scalo} J.~M.,  1979, ApJS, 41, 513

\bibitem[\protect\citeauthoryear{{Moraux}, {Bouvier} \& {Stauffer}}{{Moraux}
  et~al.}{2001}]{moraux01}
{Moraux} E.,  {Bouvier} J.,    {Stauffer} J.~R.,  2001, A\&A, 367, 211

\bibitem[\protect\citeauthoryear{{Moraux}, {Bouvier}, {Stauffer} \&
  {Cuillandre}}{{Moraux} et~al.}{2003}]{moraux03}
{Moraux} E.,  {Bouvier} J.,  {Stauffer} J.~R.,    {Cuillandre} J.-C.,  2003,
  A\&A, 400, 891

\bibitem[\protect\citeauthoryear{{Muench}, {Lada}, {Lada} \& {Alves}}{{Muench}
  et~al.}{2002}]{muench02}
{Muench} A.~A.,  {Lada} E.~A.,  {Lada} C.~J.,    {Alves} J.,  2002, ApJ, 573,
  366

\bibitem[\protect\citeauthoryear{{O'dell}, {Hendry} \& {Collier
  Cameron}}{{O'dell} et~al.}{1994}]{Odell94}
{O'dell} M.~A.,  {Hendry} M.~A.,    {Collier Cameron} A.,  1994, MNRAS, 268,
  181

\bibitem[\protect\citeauthoryear{{Offner}, {Klein}, {McKee} \&
  {Krumholz}}{{Offner} et~al.}{2009}]{offner09}
{Offner} S.~S.~R.,  {Klein} R.~I.,  {McKee} C.~F.,    {Krumholz} M.~R.,  2009,
  ApJ, 703, 131

\bibitem[\protect\citeauthoryear{{Pinfield}, {Hodgkin}, {Jameson}, {Cossburn},
  {Hambly} \& {Devereux}}{{Pinfield} et~al.}{2000}]{pinfield00}
{Pinfield} D.~J.,  {Hodgkin} S.~T.,  {Jameson} R.~F.,  {Cossburn} M.~R.,
  {Hambly} N.~C.,    {Devereux} N.,  2000, MNRAS, 313, 347

\bibitem[\protect\citeauthoryear{{Pinfield}, {Burningham}, {Tamura}, {Leggett},
  {Lodieu}, {Lucas}, {Mortlock},  \& {28 co-authors}}{{Pinfield}
  et~al.}{2008}]{pinfield08}
{Pinfield} D.~J., et al.\ 2008, MNRAS, 390, 304

\bibitem[\protect\citeauthoryear{{Rebolo}, {Zapatero-Osorio} \&
  {Mart\'{\i}n}}{{Rebolo} et~al.}{1995}]{rebolo95}
{Rebolo} R.,  {Zapatero-Osorio} M.~R.,    {Mart\'{\i}n} E.~L.,  1995, Nat, 377, 129

\bibitem[\protect\citeauthoryear{{Reyl{\'e}}, {Delorme}, {Willott}, {Albert},
  {Delfosse}, {Forveille}, {Artigau}, {Malo}, {Hill} \& {Doyon}}{{Reyl{\'e}}
  et~al.}{2010}]{reyle10}
{Reyl{\'e}} C., et al.\ 2010, A\&A, 522, A112

\bibitem[\protect\citeauthoryear{{Robichon}, {Arenou}, {Mermilliod} \&
  {Turon}}{{Robichon} et~al.}{1999}]{robichon99}
{Robichon} N.,  {Arenou} F.,  {Mermilliod} J.-C.,    {Turon} C.,  1999, A\&A,
  345, 471

\bibitem[\protect\citeauthoryear{{Salpeter}}{{Salpeter}}{1955}]{salpeter55}
{Salpeter} E.~E.,  1955, ApJ, 121, 161

\bibitem[\protect\citeauthoryear{{Scalo}}{{Scalo}}{1986}]{scalo86}
{Scalo} J.~M.,  1986, Fundamentals of Cosmic Physics, 11, 1

\bibitem[\protect\citeauthoryear{{Skrutskie}, {Cutri}, {Stiening}, {Weinberg},
  {Schneider}, {Carpenter} \& {25 co-authors}}{{Skrutskie}
  et~al.}{2006}]{skrutskie06}
{Skrutskie} M.~F., et al.\ 2006, AJ, 131, 1163

\bibitem[\protect\citeauthoryear{{Southworth}, {Maxted} \&
  {Smalley}}{{Southworth} et~al.}{2005}]{southworth05}
{Southworth} J.,  {Maxted} P.~F.~L.,    {Smalley} B.,  2005, A\&A, 429, 645

\bibitem[\protect\citeauthoryear{{Stauffer}, {Klemola}, {Prosser} \&
  {Probst}}{{Stauffer} et~al.}{1991}]{stauffer91}
{Stauffer} J.,  {Klemola} A.,  {Prosser} C.,    {Probst} R.,  1991, AJ, 101,
  980

\bibitem[\protect\citeauthoryear{{Stauffer}, {Hamilton} \& {Probst}}{{Stauffer}
  et~al.}{1994}]{stauffer94c}
{Stauffer} J.~R.,  {Hamilton} D.,    {Probst} R.~G.,  1994, AJ, 108, 155

\bibitem[\protect\citeauthoryear{{Stauffer}, {Hartmann}, {Fazio}, {Allen},
  {Patten}, {Lowrance}, {Hurt}, {Rebull}, {Cutri}, {Ramirez}, {Young}, {Rieke},
  {Gorlova}, {Muzerolle}, {Slesnick} \& {Skrutskie}}{{Stauffer}
  et~al.}{2007}]{stauffer07}
{Stauffer} J.~R., et al.\ 2007, ApJS, 172, 663

\bibitem[\protect\citeauthoryear{{Stauffer}, {Schild}, {Barrado y Navascues},
  {Backman}, {Angelova}, {Kirkpatrick}, {Hambly} \& {Vanzi}}{{Stauffer}
  et~al.}{1998}]{stauffer98b}
{Stauffer} J.~R.,  {Schild} R.,  {Barrado y Navascues} D.,  {Backman} D.~E.,
  {Angelova} A.~M.,  {Kirkpatrick} J.~D.,  {Hambly} N.,    {Vanzi} L.,  1998,
  ApJ, 504, 805

\bibitem[\protect\citeauthoryear{{Stauffer}, {Schultz} \&
  {Kirkpatrick}}{{Stauffer} et~al.}{1998}]{stauffer98}
{Stauffer} J.~R.,  {Schultz} G.,    {Kirkpatrick} J.~D.,  1998, ApJL, 499, 219

\bibitem[\protect\citeauthoryear{{Tej}, {Sahu}, {Chandrasekhar} \&
  {Ashok}}{{Tej} et~al.}{2002}]{tej02}
{Tej} A.,  {Sahu} K.~C.,  {Chandrasekhar} T.,    {Ashok} N.~M.,  2002, ApJ,
  578, 523

\bibitem[\protect\citeauthoryear{{Trumpler}}{{Trumpler}}{1921}]{trumpler21}
{Trumpler} R.~J.,  1921, Lick Observatory Bulletin, 10, 110

\bibitem[\protect\citeauthoryear{{Urban}, {Martel} \& {Evans} II}{{Urban}
  et~al.}{2010}]{urban10}
{Urban} A.,  {Martel} H.,    {Evans} II N.~J.,  2010, ApJ, 710, 1343

\bibitem[\protect\citeauthoryear{{van Leeuwen}}{{van
  Leeuwen}}{2009}]{vanLeeuwen09}
{van Leeuwen} F.,  2009, A\&A, 497, 209

\bibitem[\protect\citeauthoryear{{van Leeuwen}, {Alphenaar} \& {Brand}}{{van
  Leeuwen} et~al.}{1986}]{vanLeeuwen86}
{van Leeuwen} F.,  {Alphenaar} P.,    {Brand} J.,  1986, A\&AS, 65, 309

\bibitem[\protect\citeauthoryear{{Zapatero Osorio}, {B{\' e}jar},
  {Mart{\'{\i}}n}, {Rebolo}, {Barrado y Navascu{\' e}s}, {Mundt}, {Eisl{\"
  o}ffel} \& {Caballero}}{{Zapatero Osorio} et~al.}{2002}]{zapatero02b}
{Zapatero Osorio} M.~R.,  {B{\' e}jar} V.~J.~S.,  {Mart{\'{\i}}n} E.~L.,
  {Rebolo} R.,  {Barrado y Navascu{\' e}s} D.,  {Mundt} R.,  {Eisl{\" o}ffel}
  J.,    {Caballero} J.~A.,  2002, ApJ, 578, 536

\bibitem[\protect\citeauthoryear{{Zapatero Osorio}, {Rebolo} \&
  {Mart\'{\i}n}}{{Zapatero Osorio} et~al.}{1997}]{zapatero97a}
{Zapatero Osorio} M.~R.,  {Rebolo} R.,    {Mart\'{\i}n} E.~L.,  1997, A\&A,
  317, 164

\bibitem[\protect\citeauthoryear{{Zapatero Osorio}, {Rebolo}, {Mart{\'{\i}}n},
  {Hodgkin}, {Cossburn}, {Magazz{\` u}}, {Steele} \& {Jameson}}{{Zapatero
  Osorio} et~al.}{1999}]{zapatero99b}
{Zapatero Osorio} M.~R.,  {Rebolo} R.,  {Mart{\'{\i}}n} E.~L.,  {Hodgkin}
  S.~T.,  {Cossburn} M.~R.,  {Magazz{\` u}} A.,  {Steele} I.~A.,    {Jameson}
  R.~F.,  1999, A\&AS, 134, 537

\end{thebibliography}

%
%
\appendix

\section{Table of known Pleiades member candidates published 
in the literature and recovered in UKIDSS GCS DR9\@.}

\begin{table*}
  \caption{Sample of 1379 known Pleiades member candidates previously published 
in the literature and recovered in GCS DR9\@. We list the equatorial 
coordinates (J2000), GCS $ZYJHK1K2$ photometry, proper motions (mas/yr) and 
their errors, reduced chi-squared statistic of the astrometric fit for each 
source ($\chi^{2}$ value), membership probabilities, and names from the 
literature. Note that the 312 sources without membership probabilities are
divided into four groups: 190 non members (NM) detected in $ZYJHK$ but not
satisfying our photometric and astrometric criteria, 42 objects without $Z+Y$
photometry (no$ZY$), 74 objects without $Z$ only (no$Z$), and 6 sources 
without $Y$ only (no$Y$).
Pleiades member candidates are ordered by increasing Right Ascension.
This table is available electronically in the online version of the journal.
}
  \label{tab_Pleiades:early_DR9}
  \begin{tabular}{c c c c c c c c c c c c l}
  \hline
R.A.\ & Dec.\  &  $Z$  &  $Y$  &  $J$  &  $H$  & $K$1 & $K$2 & $\mu_{\alpha}cos\delta$\,$\pm$\,err & $\mu_{\delta}$\,$\pm$\,err & $\chi^{2}$ & Prob & Name \cr
 \hline
03 27 54.26 & +24 56 10.9 & 13.808 & 13.400 & 12.820 & 12.255 & 12.005 & 99.999 &    16.62$\pm$6.95 &   $-$43.60$\pm$6.95 &  0.47 & 0.93 & DH003   \cr
 03 29 58.76 & +23 22 18.3 & 13.640 & 13.198 & 12.672 & 12.244 & 11.843 & 11.851 &    21.18$\pm$3.41 &   $-$38.83$\pm$3.41 &  0.47 & 0.81 & DH009  \cr
 \ldots{}   & \ldots{}   & \ldots{} & \ldots{} & \ldots{} & \ldots{} & \ldots{} & \ldots{}  & \ldots{}  & \ldots{} & \ldots{} & \ldots{} & \ldots{} \cr
 04 05 13.75 & +24 08 42.7 & 14.983 & 14.471 & 13.846 & 13.261 & 12.933 & 12.917 &    19.77$\pm$3.38 &   $-$41.50$\pm$3.38 &  0.54 & 0.94 & DH915  \cr
04 06 29.99 & +22 33 43.6 & 14.376 & 13.856 & 13.201 & 12.623 & 12.273 & 12.269 &    14.83$\pm$5.07 &   $-$33.96$\pm$5.07 &  0.55 & 0.14 & DH2004\_916  \cr
 \hline
\end{tabular}
\end{table*}

\section{Table of previously-known Pleiades member candidates not recovered in the UKIDSS GCS DR9}

\begin{table*}
  \caption{Coordinates (J2000) and names of 544 previously-known Pleiades 
member candidates not recovered in the GCS DR9\@.
Pleiades candidates are ordered by increasing right ascension.
This table is available electronically in the online version of the journal.
}
  \label{tab_Pleiades:known_NOT_in_GCSDR9}
  \begin{tabular}{c c c}
  \hline
R.A.\ & Dec.\  &  Old Names \cr
 \hline
03:27:42.06 & $+$23:48:13.3 & PELS121 \cr
03:28:01.56 & $+$23:04:42.6 & DH004 \cr
\ldots{}  & \ldots{} & \ldots{} \cr
04:05:09.44 & $+$23:28:59.0 & DH913 \cr
04:05:13.72 & $+$22:18:19.0 & DH914 \cr
 \hline
\end{tabular}
\end{table*}

\section{Table of new Pleiades member candidates identified in the UKIDSS GCS DR9}

\begin{table*}
  \caption{Coordinates (J2000), near-infrared ($ZYJHK$1$K$2) photometry with the
error bars, proper motions with errors for all new Pleiades member candidates 
identified in the UKIDSS GCS DR9 with the probabilistic and standard selection 
methods. The penultimate column gives the membership probability if the object
was selected with the probabilistic method. The last column lists the source (S) 
of the object: ``1'' means that the object was identified in the probabilistic
approach, ``2'' means that the object was identified with method \#2, ``12''
means that the candidate is common to both selection method, and ``3'' means
that this is a faint $YJHK$ or $JHK$ candidate.
Pleiades member candidates are ordered by increasing Right Ascension.
This table is available electronically in the online version of the 
journal.
}
  \label{tab_Pleiades:new_members}
  \begin{tabular}{@{\hspace{0mm}}c @{\hspace{1mm}}c @{\hspace{2mm}}c @{\hspace{2mm}}c @{\hspace{2mm}}c @{\hspace{2mm}}c @{\hspace{2mm}}c @{\hspace{2mm}}c @{\hspace{2mm}}c @{\hspace{1mm}}c @{\hspace{2mm}}c @{\hspace{2mm}}c@{\hspace{0mm}}}
  \hline
R.A.\ & Dec.\  &  $Z$  &  $Y$  &  $J$  &  $H$  & $K$1 & $K$2 & $\mu_{\alpha}cos\delta$ & $\mu_{\delta}$ & Prob & S \\
 \hline
03 24 59.74 & $+$25 34 04.5 & 16.878$\pm$ 0.009 & 16.243$\pm$ 0.007 & 15.582$\pm$ 0.007 & 14.987$\pm$ 0.010 & 14.606$\pm$ 0.005 & 99.999$\pm$99.999 &    38.59$\pm$7.21 &   $-$49.57$\pm$7.21 &  ---  &  2 \cr
03 25 30.92 & $+$24 51 39.9 & 20.316$\pm$ 0.126 & 19.478$\pm$ 0.096 & 18.600$\pm$ 0.079 & 18.210$\pm$ 0.160 & 18.170$\pm$ 0.120 & 99.999$\pm$99.999 &    10.72$\pm$31.33 &   $-$45.76$\pm$31.33 & 0.60 &  1 \cr
\ldots{}  & \ldots{} & \ldots{} & \ldots{} & \ldots{} & \ldots{} & \ldots{} & \ldots{} & \ldots{}  & \ldots{} \cr
04 10 54.54 & $+$26 01 42.4 & 19.642$\pm$ 0.091 & 99.999$\pm$99.999 & 17.748$\pm$ 0.036 & 17.002$\pm$ 0.027 & 16.417$\pm$ 0.022 & 99.999$\pm$99.999 &    17.38$\pm$5.99 &   $-$47.11$\pm$5.99 & 0.66 &  1 \cr
04 11 03.84 & $+$23 15 48.9 & 17.766$\pm$ 0.017 & 17.254$\pm$ 0.014 & 16.599$\pm$ 0.014 & 15.986$\pm$ 0.017 & 15.599$\pm$ 0.021 & 15.623$\pm$ 0.012 &    18.36$\pm$6.81 &   $-$46.34$\pm$6.81 & 0.61 &  1 \cr
 \hline
\end{tabular}
\end{table*}

\section{Table of substellar multiple system candidates in the Pleiades}

\begin{table*}
  \caption{Coordinates (J2000), near-infrared ($ZYJHK$1$K$2) photometry, and proper 
motions (in mas/yr) for substellar multiple system candidates identified 
photometrically in the Pleiades cluster
}
  \label{tab_Pleiades:binary_candidates}
  \begin{tabular}{c c c c c c c c c c c}
  \hline
R.A.\ & Dec.\  &  $Z$  &  $Y$  &  $J$  &  $H$  & $K$1 & $K$2 & $\mu_{\alpha}cos\delta$ & $\mu_{\delta}$ \\
 \hline
03 25 38.73 & +22 57 39.8 & 15.890 & 15.269 & 14.601 & 13.974 & 13.618 & 13.628 &    24.45$\pm$5.11 &   $-$47.02$\pm$5.11 \cr
03 31 20.71 & +25 57 33.6 & 16.847 & 16.068 & 15.309 & 14.745 & 14.321 & 14.328 &    21.73$\pm$3.39 &   $-$38.42$\pm$3.39 \cr
03 32 11.55 & +21 27 55.7 & 16.385 & 15.652 & 14.906 & 14.339 & 13.949 & 13.920 &    13.64$\pm$3.89 &   $-$40.00$\pm$3.89 \cr
03 33 49.22 & +19 59 52.0 & 18.539 & 18.104 & 16.453 & 15.767 & 15.433 & 15.492 &     1.67$\pm$6.44 &   $-$40.37$\pm$6.44 \cr
03 34 38.61 & +24 51 28.5 & 17.097 & 16.247 & 15.459 & 14.920 & 14.445 & 14.466 &    13.39$\pm$2.69 &   $-$37.21$\pm$2.69 \cr
03 36 01.95 & +27 11 04.7 & 16.248 & 15.492 & 14.774 & 14.234 & 13.810 & 13.782 &    15.79$\pm$3.79 &   $-$42.70$\pm$3.79 \cr
03 36 03.85 & +22 52 03.5 & 18.072 & 16.873 & 15.913 & 15.240 & 14.679 & 14.679 &    22.46$\pm$3.15 &   $-$46.19$\pm$3.15 \cr
03 36 53.30 & +26 34 27.9 & 17.046 & 16.171 & 15.379 & 14.804 & 14.355 & 14.353 &    15.54$\pm$3.33 &   $-$34.04$\pm$3.33 \cr
03 40 11.98 & +21 48 31.8 & 16.578 & 15.884 & 15.169 & 14.580 & 14.187 & 14.178 &    24.16$\pm$2.54 &   $-$40.67$\pm$2.54 \cr
03 40 45.16 & +27 50 40.4 & 17.074 & 16.218 & 15.404 & 14.816 & 14.321 & 14.333 &    12.37$\pm$3.34 &   $-$40.37$\pm$3.34 \cr
03 40 53.66 & +28 21 11.5 & 18.880 & 17.570 & 16.490 & 15.852 & 15.194 & 15.199 &    13.91$\pm$4.06 &   $-$31.57$\pm$4.06 \cr
03 41 40.91 & +25 54 24.1 & 16.893 & 16.001 & 15.180 & 14.574 & 14.122 & 14.125 &    16.94$\pm$2.26 &   $-$42.13$\pm$2.26 \cr
03 41 42.41 & +23 54 57.1 & 16.171 & 15.464 & 14.709 & 14.124 & 13.686 & 99.999 &    14.07$\pm$2.31 &   $-$47.37$\pm$2.31 \cr
03 41 54.16 & +23 05 04.7 & 17.349 & 16.376 & 15.522 & 14.975 & 14.415 & 14.418 &    18.19$\pm$2.30 &   $-$44.74$\pm$2.30 \cr
03 43 34.49 & +25 57 30.6 & 16.571 & 15.727 & 14.909 & 14.359 & 13.909 & 13.901 &    20.92$\pm$2.25 &   $-$47.72$\pm$2.25 \cr
03 44 14.65 & +23 49 40.0 & 15.892 & 15.245 & 14.547 & 13.957 & 13.546 & 99.999 &    14.46$\pm$2.31 &   $-$40.15$\pm$2.31 \cr
03 44 35.16 & +25 13 42.8 & 17.656 & 16.584 & 15.662 & 14.985 & 14.448 & 14.460 &    19.33$\pm$2.26 &   $-$44.97$\pm$2.26 \cr
03 44 35.90 & +23 34 41.9 & 16.307 & 15.672 & 14.985 & 14.376 & 13.990 & 13.985 &    16.94$\pm$2.23 &   $-$44.39$\pm$2.23 \cr
03 45 09.46 & +23 58 44.7 & 16.974 & 16.250 & 15.438 & 14.872 & 14.424 & 14.410 &    16.07$\pm$2.25 &   $-$42.23$\pm$2.25 \cr
03 45 31.37 & +24 52 47.4 & 17.332 & 16.330 & 15.465 & 14.839 & 14.354 & 14.326 &    16.69$\pm$2.24 &   $-$40.30$\pm$2.24 \cr
03 45 37.76 & +23 43 50.1 & 16.240 & 15.456 & 14.715 & 14.172 & 13.756 & 13.742 &    20.91$\pm$2.23 &   $-$45.45$\pm$2.23 \cr
03 45 41.27 & +23 54 09.7 & 17.166 & 16.189 & 15.360 & 14.782 & 14.305 & 14.309 &    17.46$\pm$2.24 &   $-$44.47$\pm$2.24 \cr
03 45 50.66 & +24 09 03.5 & 17.478 & 16.582 & 15.705 & 15.095 & 14.580 & 14.560 &    16.01$\pm$2.26 &   $-$40.58$\pm$2.26 \cr
03 46 05.11 & +23 45 34.9 & 17.229 & 16.347 & 15.522 & 14.851 & 14.385 & 14.404 &    15.88$\pm$2.25 &   $-$39.21$\pm$2.25 \cr
03 46 15.11 & +26 46 48.8 & 16.624 & 15.838 & 15.032 & 14.495 & 14.037 & 14.088 &    20.73$\pm$2.97 &   $-$42.79$\pm$2.97 \cr
03 46 20.27 & +23 58 18.9 & 19.259 & 18.174 & 17.034 & 16.269 & 15.650 & 15.585 &    12.97$\pm$2.40 &   $-$35.00$\pm$2.40 \cr
03 46 22.25 & +23 52 26.6 & 17.120 & 16.323 & 15.518 & 14.917 & 14.474 & 14.472 &    17.65$\pm$2.25 &   $-$38.04$\pm$2.25 \cr
03 46 26.09 & +24 05 09.5 & 16.810 & 15.967 & 15.160 & 14.584 & 14.118 & 14.098 &    19.41$\pm$2.24 &   $-$38.71$\pm$2.24 \cr
03 46 27.10 & +21 48 22.6 & 19.797 & 18.650 & 17.374 & 16.564 & 15.848 & 15.925 &    20.95$\pm$2.98 &   $-$48.67$\pm$2.98 \cr
03 47 11.79 & +24 13 31.3 & 16.305 & 15.554 & 14.792 & 14.261 & 13.841 & 13.850 &    16.88$\pm$2.23 &   $-$41.27$\pm$2.23 \cr
03 47 20.48 & +19 54 25.5 & 17.011 & 16.054 & 15.210 & 14.615 & 14.152 & 14.140 &    28.11$\pm$5.10 &   $-$39.39$\pm$5.10 \cr
03 48 04.67 & +23 39 30.1 & 17.014 & 16.054 & 15.283 & 14.704 & 14.256 & 14.238 &    16.07$\pm$2.24 &   $-$44.27$\pm$2.24 \cr
03 48 31.53 & +24 34 37.2 & 19.218 & 17.883 & 16.715 & 15.977 & 15.357 & 15.312 &    11.92$\pm$2.37 &   $-$46.68$\pm$2.37 \cr
03 48 35.20 & +22 53 42.1 & 16.176 & 15.452 & 14.745 & 14.185 & 13.770 & 13.772 &    15.22$\pm$2.15 &   $-$45.62$\pm$2.15 \cr
03 48 50.45 & +22 44 29.8 & 16.562 & 15.825 & 15.098 & 14.531 & 14.141 & 14.143 &    15.64$\pm$2.16 &   $-$42.06$\pm$2.16 \cr
03 48 57.41 & +23 13 59.1 & 16.835 & 16.033 & 15.222 & 14.673 & 14.194 & 14.194 &    14.93$\pm$2.17 &   $-$36.62$\pm$2.17 \cr
03 50 52.17 & +23 27 11.2 & 17.690 & 16.638 & 15.700 & 15.060 & 14.505 & 14.555 &    19.89$\pm$2.21 &   $-$41.33$\pm$2.21 \cr
03 51 38.96 & +24 30 44.8 & 18.711 & 17.407 & 16.400 & 15.718 & 15.168 & 15.122 &    23.01$\pm$2.34 &   $-$40.47$\pm$2.34 \cr
03 53 55.13 & +23 23 36.1 & 16.947 & 16.027 & 15.172 & 14.569 & 14.088 & 14.081 &    19.17$\pm$2.28 &   $-$44.72$\pm$2.28 \cr
03 55 27.06 & +25 14 45.8 & 16.118 & 15.402 & 14.649 & 14.071 & 13.671 & 13.650 &    15.24$\pm$2.24 &   $-$39.66$\pm$2.24 \cr
03 56 52.31 & +25 10 05.1 & 16.146 & 15.491 & 14.756 & 14.192 & 13.771 & 13.801 &    16.66$\pm$2.50 &   $-$38.22$\pm$2.50 \cr
03 58 00.62 & +21 18 20.8 & 17.399 & 16.380 & 15.545 & 14.931 & 14.445 & 14.423 &    24.81$\pm$3.43 &   $-$32.13$\pm$3.43 \cr
03 58 17.43 & +22 11 52.7 & 16.259 & 15.503 & 14.768 & 14.193 & 13.787 & 13.778 &    20.64$\pm$2.89 &   $-$34.95$\pm$2.89 \cr
03 59 59.85 & +25 08 53.6 & 16.368 & 15.695 & 15.004 & 14.433 & 14.029 & 14.075 &    14.87$\pm$2.51 &   $-$35.69$\pm$2.51 \cr
04 00 03.21 & +22 24 46.0 & 16.534 & 15.875 & 15.128 & 14.550 & 14.132 & 14.141 &    15.92$\pm$2.87 &   $-$33.98$\pm$2.87 \cr
04 00 08.16 & +22 32 01.1 & 16.449 & 15.820 & 15.102 & 14.503 & 14.098 & 14.104 &    16.31$\pm$2.88 &   $-$39.63$\pm$2.88 \cr
04 00 50.52 & +23 43 52.9 & 16.184 & 15.380 & 14.647 & 14.089 & 13.660 & 99.999 &    24.52$\pm$3.56 &   $-$40.90$\pm$3.56 \cr
04 01 28.43 & +23 30 59.6 & 16.318 & 15.539 & 14.772 & 14.202 & 13.788 & 13.788 &    24.55$\pm$3.37 &   $-$39.20$\pm$3.37 \cr
04 01 39.83 & +22 47 53.7 & 16.669 & 15.899 & 15.153 & 14.575 & 14.169 & 14.157 &    21.16$\pm$3.39 &   $-$42.62$\pm$3.39 \cr
04 01 50.95 & +22 59 15.5 & 16.363 & 15.631 & 14.906 & 14.311 & 13.912 & 13.906 &    20.72$\pm$3.38 &   $-$38.98$\pm$3.38 \cr
04 09 17.80 & +26 03 31.2 & 17.120 & 16.480 & 15.759 & 15.003 & 14.603 & 99.999 &     6.56$\pm$4.49 &   $-$33.73$\pm$4.49 \cr
 \hline
\end{tabular}
\end{table*}

\end{document}